\pdfoutput=1

\documentclass[acmsmall,screen, nonacm,natbib=false]{acmart}

\newboolean{appendixincluded}
\setboolean{appendixincluded}{true}

\newcommand{\appendixref}[2]{\ifthenelse{\boolean{appendixincluded}}{Appendix~\ref{#1}}{the appendix~\cite[{#2}]{appendix}}}

\usepackage[capitalize]{cleveref}

\crefname{line}{Line}{Lines}
\crefname{rule}{rule}{rules}
\Crefname{rule}{Rule}{Rules}
\crefrangelabelformat{line}{#3#1#4--#5#2#6}
\usepackage{thmtools}
\usepackage{thm-restate}

\usepackage{wrapfig}
\usepackage[inline,shortlabels]{enumitem}
\usepackage{mathpartir} 
\usepackage{style/comments}
\usepackage{style/notation}
\usepackage{style/iris}
\usepackage{style/heaplang}

\RequirePackage[
  datamodel=acmdatamodel,
  style=acmnumeric,
  ]{biblatex}

\newcommand{\thelogic}{\textsf{Amaryllis}}

\newcommand{\Bluebell}{\textsc{Bluebell}}
\newcommand{\pcOL}{\textsc{pcOL}}
\newcommand{\BaSL}{\textsc{BaSL}}

\begin{document}

\title[First Steps Towards Probabilistic Iris: Harmonizing Independence, Conditioning, and Dynamic Heap Allocation]{First Steps Towards Probabilistic Iris: Harmonizing Independence, Conditioning, and Dynamic Heap Allocation}

\author[J. Lohse]{Janine Lohse}
\affiliation{
  \institution{MPI-SWS}            \country{Germany}
}
\email{jlohse@mpi-sws.org}          

\author[T. Rohde]{Tim Rohde}
\affiliation{
  \institution{MPI-SWS and Saarland University}            \country{Germany}
}
\email{timrohde@mpi-sws.org}          

\author[J. Xin]{Jimmy Xin}
\affiliation{
  \institution{MPI-SWS}            \country{Germany}
}
\email{jimmyxin@mpi-sws.org}          

\author[N. Mück]{Niklas Mück}
\affiliation{
  \institution{MPI-SWS}            \country{Germany}
}
\email{mueck@mpi-sws.org}          

\author[I. Kuhn]{Iona Kuhn}
\affiliation{
  \institution{CISPA Helmholtz Center for Information Security}            \country{Germany}
}
\email{iona.kuhn@cispa.de}          

\author[D. Dreyer]{Derek Dreyer}
\affiliation{
  \institution{MPI-SWS}
\country{Germany}
}
\email{dreyer@mpi-sws.org}

\author[D. Garg]{Deepak Garg}
\affiliation{
  \institution{MPI-SWS}
\country{Germany}
}
\email{dg@mpi-sws.org}

\author[E. D'Osualdo]{Emanuele D'Osualdo}
\authornote{Lead senior author.}
\affiliation{
  \institution{University of Konstanz}
\country{Germany}
}
\email{emanuele.dosualdo@uni-konstanz.de}

\authorsaddresses{}

\begin{abstract}
There has recently been exciting progress in the realm of \emph{probabilistic separation logics}.
An important subclass of these---including PSL, Lilac, \Bluebell, and \pcOL---are \emph{distributional probabilistic logics} (or DPLs, for short), meaning that they provide primitive Hoare-style assertions about probability distributions on the program state, along with powerful modularity principles like \emph{independence} and \emph{conditioning}.
However, none of these logics support reasoning about dynamically allocated memory (\ie pointers into a heap), let alone the more sophisticated resource algebra-based ghost state of modern separation logics like Iris.
We argue that this is due to a fundamental obstacle: 
since the shape of memory (and identity of memory locations) may differ under different random outcomes,
it is unclear how pointer ownership can be harmonized with probabilistic independence and conditioning.
Furthermore, none of the existing DPLs that support both independence and conditioning have, to our knowledge, been mechanized in a proof assistant.

In this paper, we take substantial first steps towards a marriage of DPLs and modern separation logics like Iris, in the form of \textbf{\thelogic{}}.
\thelogic{} is the first DPL to support independence and conditional reasoning while also handling dynamic memory allocation.
To overcome the aforementioned obstacle, we propose
a new \emph{indexed valuation}-style model of probabilistic assertions,
whereby ownership and composition of standard Iris-style resources (\eg heaps) can be promoted to corresponding notions of ownership and composition at the level of distributions by interpreting them \emph{per random outcome}.
We then show how to adapt the central Iris notions of \emph{frame-preserving update}, \emph{authoritative resource algebras}, and the \emph{weakest precondition modality} to be sound for probabilistic reasoning and validate dynamic allocation.
Finally, we have mechanized all our results in the Rocq proof assistant and developed an Iris-based proof mode for conducting proofs within \thelogic.

 \end{abstract}

\maketitle

\section{Introduction}
\label{sec:intro}

The past few years have seen a flurry of work on program logics for reasoning modularly about probabilistic programs.
Roughly speaking, these probabilistic logics can be divided into two types, which may be understood as ``lifting-based'' vs.\ ``distributional''.
\emph{Lifting-based probabilistic logics} (or LPLs)~\cite{eris,coneris,tachis,clutch,approxis,wpexp} (1) encode various specific probabilistic properties of interest (\eg error bounds, expected time complexity, etc.) using custom forms of ghost state, then (2) derive rules for reasoning about them within the framework of standard Hoare or separation logics~\cite{sl-reynolds,sl-ohearn} (\eg Iris~\cite{iris1,iris}), and finally (3) provide an adequacy theorem that lifts specifications in terms of the ghost state assertions to valid probabilistic statements at the meta-level.
\emph{Distributional probabilistic logics} (or DPLs)~\cite{ellora,psl,lilac,bluebell,pcol} instead provide native support for program predicates describing \emph{distributions} over program states, as well as logical connectives describing \emph{independence} and \emph{conditioning} of distributions, neither of which are directly expressible in existing LPLs.

Thus far, the LPLs have exhibited the advantage that, since they are built from (in some sense) more standard parts (\eg defining some appropriate form of ghost state and applying the standard Iris methodology), they are easier to construct and soundly apply in the context of more fully-featured programming languages (\eg with higher-order state).
In contrast, the DPLs have the potential advantage of being more general and expressive, supporting a wider range of modular probabilistic reasoning principles in one logic.

However---notably---none of the existing DPLs support reasoning about \emph{dynamically allocated memory} (\ie pointers into a heap), let alone the more sophisticated resource algebra-based ghost state of modern separation logics like Iris.
This may come as a surprise, given that dynamically allocated pointers were the \emph{raison d'être} of the original version of separation logic~\cite{sl-ohearn}.
But there is good reason for it: multiple, fundamental technical challenges must be overcome in order to harmonize DPLs' support for independence and conditioning---implemented using a non-standard model of separating conjunction---with the \emph{resource}-based (\eg heap) model of separation in traditional separation logics.

As a long-term goal, we aim to marry together the benefits of LPLs and DPLs in the form of Probabilistic Iris: an adaptation of the Iris framework that would support rich programming language features and modern separation logic features \emph{alongside} primitive assertions about probability distributions.
In this paper, we take a substantial first step towards that goal with \textbf{\thelogic}, the first DPL to support reasoning about Iris-style resources, including dynamically allocated memory (heaps).
Though we use heaps as a motivation, \thelogic{}'s approach is not actually specific to heap resources at all: the logic is parametric in a (non-step-indexed) Iris-style resource algebra describing stateful, non-probabilistic resources, and shows how to generically lift such a resource algebra to a model of probabilistic assertions supporting independence, conditioning, and modular reasoning about state.
This generic lifting requires a fundamental rethink of various aspects of existing DPL models.
To build added confidence in the resulting constructions,
we have fully mechanized \thelogic{} in the proof assistant (Rocq/Iris), including building support for interactive proofs atop the Iris Proof Mode.
To our knowledge, this constitutes the first mechanization of a DPL supporting both independence and conditional reasoning.

\textbf{Contributions.}
In summary, we make the following contributions:
\begin{enumerate}
\item
We observe that existing DPL models---which interpret assertions as predicates on distributions over program states---run into a major road block when we incorporate dynamic memory allocation because they do not account for the possibility that the shape of memory (and identity of memory locations) may differ under different random outcomes.
\item
We resolve this problem by proposing an \emph{indexed-valuation}-style model for DPLs in which probabilistic resources comprise two parts: (1) a probability distribution on \emph{outcomes} (\ie results of coin flips), and (2) a mapping from outcomes to separation logic resources (\eg heaps).
In this model, separating conjunction is defined \emph{per random outcome}, rather than requiring disjointness of heaps across all outcomes as prior models do.
Consequently, it allows different random outcomes to result in differently shaped heaps, enabling a clean accounting of dynamic heap allocation.
\item
On top of this probabilistic resource model, we build a program logic, \thelogic, which supports key reasoning principles of prior DPLs---notably, independence and conditional reasoning---while also supporting dynamic allocation.
\item
We show for the first time how to adapt central concepts in Iris---such as ``frame-preserving update'' and ``the authoritative resource algebra'', which are used in turn to derive a weakest-precondition modality---so that they remain sound under probabilistic (and conditional) reasoning.
\item
We demonstrate the effectiveness of \thelogic\ on a range of illustrative examples.
\item
We have mechanized all results of this paper in the Rocq proof assistant, and also developed an Iris-based proof mode for conveniently conducting proofs within \thelogic.\footnote{Some of the Rocq proofs were implemented with the assistance of Codex and Claude Code, but for all theorem statements and definitions mentioned in the paper, we formalized the specifications ourselves in Rocq.}
\end{enumerate}

\textbf{Non-goals, limitations, and future work.}
In order to focus attention on the core problem of handling dynamic heap allocation and Iris-style resources, we place several restrictions on the probabilistic setting and the types of programs we consider.
Specifically, we restrict attention to terminating programs (as Lilac~\cite{lilac} and \Bluebell~\cite{bluebell} do) and discrete probability distributions (as \Bluebell~\cite{bluebell} and \pcOL~\cite{pcol} do).
We also do not attempt to incorporate Iris's mechanisms for step-indexing~\cite{Appel2001PPC,Ahmed2004Thesis}, higher-order ghost state~\cite{Jung2016HigherOrder}, and concurrency/invariants~\cite{iris1}.
Lastly, for simplicity, we only consider distributions with finite support.
In future work, we expect that it should be possible to generalize \thelogic\ to handle countably infinite support and almost-sure termination of unbounded loops, but that some of \thelogic's proof rules for conditional reasoning will not apply to continuous distributions.

 \section{Key Ideas}
\label{sec:key_ideas}

In this section we give an overview of the main contributions of \thelogic.

\subsection{State-of-the-Art Reasoning in DPLs}
\label{sec:key:state_of_art}

The main judgments of distributional probabilistic logics (DPLs) are Hoare triples of the form
$ \hoare{P}{e}{\Ret V.Q(V)} $.
Here,~$e$ is an expression in a probabilistic language, the semantics of which can be seen as
a function taking a distribution over the initial state and returning
a joint distribution of final state and return value.
The \emph{precondition}~$P$ is an assertion constraining the input distribution,
while the \emph{postcondition}~$Q$ constrains the output distribution.
A \emph{random~$T$ variable} (for some type~$T$) is a function\footnote{Random variables are usually required to be measurable in some ambient probability space. To avoid confusion we will explicitly state, when relevant,
if we assume them to be measurable, and in which space they are.}
from the outcomes of a distribution to~$T$.
The postcondition is parametric on~$V$, which is a random value variable
representing the return value.
As an example,
$\hoare{\distAs{X}{\Ber{\onehalf}}}{e}{\Ret V.\distAs{V}{\Ber{\onehalf}}}$
states that the program~$e$ returns a value distributed as a fair coin
if~$X$ is initially distributed as a fair coin.

\thelogic\ builds on previous work, starting from PSL~\cite{psl}
and further developed in Lilac~\cite{lilac}, \Bluebell~\cite{bluebell}, and \pcOL~\cite{pcol}.
The success of these logics lies in modular reasoning, which is enabled
by two crucial ideas:
\begin{enumerate*}
  \item \emph{probabilistic independence} as separation, and
  \item \emph{conditioning} as a modality.
\end{enumerate*}

\paragraph{Independence as separation}
PSL introduced the idea that separating conjunction $P * Q$ can model
probabilistic independence in a logic of assertions over distributions.
For example, an assertion $\distAs{X}{\Ber{\onehalf}} * \distAs{Y}{\Ber{\onehalf}}$ does not just state that both $X$ and $Y$ are distributed as
fair coins, but also that they are \emph{independent}:
knowing the outcome of one does not reveal any information about the outcome of the other. More precisely, the probability of seeing both $v$ in $X$ and $w$ in $Y$ is the product of the two constituent probabilities.

Internalizing independence as separation introduces a very useful
principle of \emph{locality}, expressed as the Frame rule of Separation Logic:
\begin{proofrule}
  \infer*[lab=Frame]{
    \hoare{P}{e}{\Ret V.Q(V)}
  }{
    \hoare{P*R}{e}{\Ret V.Q(V)*R}
  }
  \label{rule:frame}
\end{proofrule}
Suppose, for instance, that we have
$ \hoare{\EMP}{e}{\Ret X. \distAs{X}{\Ber{\onehalf}}} $;
then by the \ref{rule:frame} rule we can prove
$ \hoare{\distAs{Y}{\Ber{\onehalf}}}{e}{\Ret X. \distAs{X}{\Ber{\onehalf}}*\distAs{Y}{\Ber{\onehalf}}} $
without having to verify~$e$ again from the larger precondition.
Since generating/consuming independent variables is commonplace in
probabilistic programs, framing turns out to be extremely valuable.

\paragraph{Conditioning as a modality}
However, framing alone is not enough to achieve modularity in the probabilistic
setting.
Imagine we are provided with the triple
\begin{equation}
\forall b \in \Bool.
  \hoare {\sure{\loc\mapsto b}}{f(\loc)}{\sure{\loc\mapsto \neg b}}
  \label{triple:det-neg}
\end{equation}
where $\sure{\hole}$ states that the predicate inside is true with probability~1.
Here the contents of location~$\loc$ are assumed to be the deterministic
value~$b$.
Intuitively, we should be able to prove
that if the contents of $\loc$ are distributed as a fair coin,
written $ \loc \pmapsto \Ber{\onehalf} $,
then
\begin{equation}
  \hoare {\loc \pmapsto \Ber{\onehalf}}{f(\loc)}{\loc \pmapsto \Ber{\onehalf}}
  \label{triple:flip-id-neg}
\end{equation}
holds.
But the frame rule alone is not sufficient to derive \eqref{triple:flip-id-neg} from \eqref{triple:det-neg}---indeed, this derivation is not possible in the original PSL.
Intuitively, we need to merge the two instances of \eqref{triple:det-neg}
(for the two choices of~$b$)
into a single one by blending them with half probability each.
To represent such ``blending'' at the level of assertions,
one can define an operator $ P \condop[q] Q $ which,
roughly speaking, should hold on a distribution if it can be
seen as the sum weighted by $q\in[0,1]$ and $1-q$ of two distributions
which satisfy~$P$ and~$Q$ respectively.
With this addition we can formulate a second principle of locality---\emph{outcome locality}---in the form of the following rule:
\begin{proofrule}
  \infer*[lab={${\condop[]}$-lift}]{
    \hoare{P_1}{e}{Q_1}
    \\
    \hoare{P_2}{e}{Q_2}
  }{
    \hoare{P_1 \condop[q] P_2}{e}{Q_1 \condop[q] Q_2}
  }
  \label{rule:condop-lift}
\end{proofrule}
In our example, one can establish
$ \loc \pmapsto \Ber{\onehalf} \equiv \sure{\loc\mapsto 0} \condop \sure{\loc\mapsto 1} $ and \eqref{triple:flip-id-neg} then follows
from~\ref{rule:condop-lift}.

This kind of reasoning is an instance of a more general pattern:
\emph{reasoning under conditioning}.
Often in probabilistic proofs it is necessary to perform case analysis on
some random variable~$X$ (like the value at $\loc$ in the example).
This formally corresponds to proving something about a distribution~$\dist$,
by considering the family of \emph{conditional distributions} $\dist|_{X=v}$
(\ie the restriction of $\dist$ to the case where~$X$ is the deterministic value~$v$),
one for each possible value~$v$.

The Lilac logic~\cite{lilac} proposed to represent conditioning using
 a \emph{conditioning modality},
 which can be seen as a generalization of $\condop[q]$.
In \thelogic\ we directly adopt this idea, building on later variants
of the modality defined for discrete distributions in \cite{bluebell,pcol}.
Intuitively, \thelogic's conditioning modality
$ \cond{\dist}{x}{P(x)} $ holds if
there is
a random variable~$X$ distributed as~$\dist$ such that
for every $v\in\supp(\dist)$,
$P(v)$ holds on the distribution conditioned on $X=v$;
for example, $\sure{\loc\mapsto 0} \condop[q] \sure{\loc\mapsto 1}$ can be
expressed as $ \cond{\Ber{q}}{b}{\sure{\loc\mapsto b}} $.
The modality $ \cond{\dist}{x}{P(x)} $---formally defined in \cref{sec:cond-wsum}---leaves implicit
(\ie existentially quantified)
which variable~$X$ is being conditioned on, but one must exist;
to make~$X$ apparent, it is sufficient to write $P(x)$ such that
$ \forall x\st P(x) \entails \sure{X=x} $.

The growing body of work on probabilistic separation logics provides convincing evidence
that these principles are key to making reasoning scale.
Yet, existing logics either do not support mutable state (Lilac)
or only support a finite variable store (PSL, \Bluebell, \pcOL).
This is not by chance: as we now explain, dynamic heap allocation poses serious challenges for the techniques developed~so~far.

\subsection{The Challenge of Harmonizing Independence with Dynamic Allocation}
\label{sec:key:dynalloc}

The crux of the issue with previous models and dynamic allocation
is the difficulty of harmonizing
\emph{probabilistic independence} and \emph{disjointness of state} in a single logic.

Lilac~\cite{lilac} uses separation exclusively for independence,
and thus does not support reasoning about mutable state.
In logics that additionally support mutation, one needs to embed in the model
information about disjointness of locations.
To achieve this, both PSL~\cite{psl} and \pcOL~\cite{pcol} adopt a definition of separating conjunction in which independence and disjointness are conflated.
That is, they define a distribution of states $\dist$ to satisfy $P * Q$ only if
there are two \emph{disjoint} sets of locations $ L_1, L_2 $
such that for some $\dist_1$ distribution of memories of $L_1$ satisfying $P$
and some $\dist_2$ distribution of memories of $L_2$ satisfying~$Q$
we have $\dist = \dist_1 \pprod \dist_2$
(where $\pprod$ is the independent product).

Conflating independence and disjointness in the meaning of $P * Q$ is convenient in that it neatly divides the set of memory locations that $P$ cares about from the set of memory locations that $Q$ cares about.  
Unfortunately, however, such a neat division of concerns does not work in the presence of dynamic allocation.
To illustrate why that is, let us assume a deterministic allocator for the moment,
to show that the core issues arise equally in this simpler setting.
In the technical development, \thelogic\ only assumes a non-probabilistic
but non-deterministic allocator, obtaining an even stronger result.

What makes dealing with allocation hard is the necessary dependency
of the allocator on the current heap
(as it needs, at the very least, to know which locations are available).
Since the size and shape of the heap may depend on previous random choices made by the program,
the specific locations picked by the allocator will also depend
on previous random choices.

To see the issue concretely, consider the following program:
\begin{equation}
  \begin{aligned}
  \var{dfl} \is {}&
    \Let \wtv=(\If \flip then\Alloc{0}) in
    (\Alloc{\flip}, \Alloc{\flip})
  \end{aligned}
  \label{ex:dfl}
\end{equation}
Depending on an initial coin flip, the heap may grow by one location
before the two returned locations are allocated.
Ideally, we would like to prove a postcondition that asserts
that we end up with two disjoint heap locations storing the results of two independent coins:
\begin{equation}
  \hoare {\TRUE}{\var{dfl}}{
    \Ret(X,Y).X \pmapsto \Ber{\onehalf} * Y \pmapsto \Ber{\onehalf}
  }
  \label{triple:dfl}
\end{equation}
Unfortunately, this triple does not hold (semantically) in the model sketched above.
The reason is that it is impossible to partition the locations into~$L_1,L_2$ so
that~$X$ only gets locations from $L_1$ and~$Y$ from $L_2$,
across all executions of the program.
For instance, starting from an empty heap, and assuming locations are represented as integers starting at $\loclit0$,
if the first flip is 0 the two returned locations would be
$(\loclit0,\loclit1)$,
but would be
$(\loclit1,\loclit2)$
if the first flip were 1.
Therefore, although in any particular execution $X$ and $Y$ will be two distinct locations,
when considering the whole distribution at once,
the sets of possible locations associated with~$X$ and with~$Y$
both include $\loclit1$.

\Bluebell\ proposed a fractional-permission-based way of decoupling disjointness from independence; their solution, however, is specific to static variable stores, and does not obviously generalize to heaps with dynamic allocation.
Moreover, the \Bluebell\ logic only supports the \ref{rule:condop-lift} rule
with side conditions which disable the use of frame around it,
hindering modularity.  (See \cref{sec:relwork})

\subsection{A Model of Per-Outcome Disjointness}
\label{sec:key:model}

In Amaryllis, we take a different approach to the problem of harmonizing independence and heap separation.
For $P * Q$, rather than insisting that the set of locations associated with $P$ (across all executions of the program) is disjoint from the set of locations associated with $Q$, we instead merely insist that $P$ and $Q$ assert disjoint ownership \emph{per random outcome of the program} (\ie under any given outcome of the coin flips made by the program).
Returning for instance to the problematic $\mathit{dfl}$ example given in the previous section, observe that it is perfectly fine to assert ownership of $X$ and $Y$ separately if one interprets disjointness per random outcome, since under any particular outcome the location chosen to represent $X$ and the location chosen to represent $Y$ are distinct.

To realize this idea, we propose a new model of probabilistic
separation logic which is parametric on a standard resource algebra.
Specifically, let us assume we are given a standard Iris-style resource algebra~$\monoid$ formalizing the desired notion of (non-probabilistic) resource ownership and composition.
For example, in the case of heaps, the resources would be finite maps~$h$ from locations to values;
heap composition $h_1 \mop h_2$ would be defined as disjoint union~$h_1 \dunion h_2$ if the two maps have disjoint domains, and undefined otherwise.
Our goal is then to show how to build a resource algebra representing \emph{distributions}
of elements of~$\monoid$, in such a way that we can model assertions about
  distributions of random variables,
  probabilistic conditional independence, and
  ownership of memory in one logic.

The crucial question is: what is the correct way of lifting the operation $\mop$
to an operation~$\lmop$ on \emph{distributions} of resources~$\Dist(\monoid)$?
A natural choice is to make $\dist_1 \lmop \dist_2$ defined only if
the composition $r_1 \mop r_2$ is defined for
\emph{every} $r_1\in\supp(\dist_1)$ and $r_2\in\supp(\dist_2)$
(\ie for every pair of resources that have non-zero probability).
This choice would give rise to the model of PSL and \pcOL, which---as we have explained already---is not what we want because it forces consideration of pairs of resources that do not arise in the same random outcome of the program.

This begs the question: how can we formalize the idea of only composing resources \emph{per random outcome}?
Doing so is not straightforward if we work directly with the domain $\Dist(\monoid)$ since this domain does not provide a way of talking about different random outcomes explicitly.

In \thelogic, we therefore move to a different representation of distributions
of resources, which is a variant of so-called \emph{indexed valuations}
\cite{VaraccaW06}, constructed as follows.
We assume an arbitrary countable set of
\emph{random choice identifiers}~$\rid\in\Rid$.
We then fix the outcome space to $ \Outcomes = \Rid \to \Bool $ of maps
from random choice identifiers to booleans:\footnote{
  Although~$\Outcomes$ is uncountable,
  the probability spaces we consider
  only contain information about a finite set~$R \subs \Rid$.
  We make this precise in \cref{sec:model}.
}
these abstractly represent the coin tosses performed by the program so far.
At a high level, we define a probabilistic resource to be a pair $(\psp, R)$
where
  $\psp$ is a probability space over outcomes $\Outcomes$, and
  $\rvRes \from \Outcomes \to \monoid$ is a \emph{random resource variable}
  (which is \emph{not} required to be measurable in $\psp$).
Thanks to the fact that we have an abstract representation of
the outcomes of random choices as~$\outc\in\Outcomes$,
all our resources are now indexed in~$\rvRes$ by the specific random branch
they belong to.
This allows us to define $(\psp_1, \rvRes_1) \lmop (\psp_2, \rvRes_2)$ as the pair
$ \bigl( \psp_1 \iprod \psp_2, (\fun\outc.\rvRes_1(\outc) \mop \rvRes_2(\outc)) \bigr) $,
that is:
the probability spaces are composed using
the independent product~$\iprod$ of Lilac,
while the stateful resources (\eg heaps) are composed \emph{outcome-wise}.

In the \thelogic\ model, we can give the desired meaning to
the triple in~\eqref{triple:dfl}.
The assertion $ X \pmapsto \Ber{\onehalf} $ is given a meaningful semantics
by letting $X \from \Outcomes \to \Loc$~be a \emph{random location variable}.
Given some $\dist \in \Dist(\Val)$,
the assertion $ X \pmapsto \dist $ holds on $(\psp,R)$ if
the function $\fun\outc. \rvRes(\outc)(X(\outc))$ is measurable in $\psp$
and is distributed as~$\dist$ in $\psp$.
In particular, $X(\rho)$ needs to be a location owned by $R(\rho)$,
for every outcome $\rho$ with non-zero probability in~$\psp$.
Then the triple in~\eqref{triple:dfl} expresses exactly that
\begin{enumerate*}[(i)]
  \item outcome-wise, the two functions~$X$ and $Y$ are different locations; and
  \item they point to values distributed as independent fair coins.
\end{enumerate*}

\subsection{Probabilistic Specifications in \thelogic}
\label{sec:key:specs}

Now that we have an appropriate model of distributions of heaps,
we want to build a program logic.
Consider a possible axiom for allocation:
\begin{proofrule}
  \infer*[lab=hoare-alloc-v]{}{
      \hoare{\TRUE} {\Alloc{v}} {\Ret L. L \mapsto v}
    }
  \label{rule:hoare-alloc-v}
\end{proofrule}
The rule uses \thelogic's ``points-to'' assertion,
which is defined in general as follows.
Given $L\from\Outcomes\to\Loc$ and $V\from\Outcomes\to\Val$,
the assertion $L \mapsto V$ holds on a resource $(\psp,R)$ just if
$L(\outc) \in \dom(R(\outc))$ and $R(\outc)(L(\outc)) = V(\outc)$
for all $\outc\in\supp(\psp)$.
That is, the assertion $L \mapsto V$ implies per-outcome ownership of the \emph{heap resource} $L(\outc) \mapsto V(\outc)$, but crucially it \emph{does not} imply ownership of the \emph{distribution} of either $L$ or $V$.
This is because
\begin{enumerate*}[(i)]
  \item in the model, $R$ is not required to be measurable in~$\psp$,
  and
  \item the assertion only constrains~$R$, not $\psp$.
\end{enumerate*}

Notice in particular how if $L \mapsto V$ were to own the distribution of~$L$, it would be disastrous.
As explained in \cref{sec:key:dynalloc}, $L$~is potentially dependent on (\ie correlated with) every past random choice.
Hence, if $L \mapsto V$ asserted something about the distribution of~$L$, the postcondition of $\Alloc(v)$ would not be independent of the program context, and the specification of $\Alloc(v)$ would not be frame-preserving.
By confining $L \mapsto V$ to describing heap and not distribution ownership, we ensure that the specification enjoys the frame rule.
The assertion $ L \pmapsto \dist $
which we used in \cref{sec:key:state_of_art}
can now be defined as shorthand for
$\exists V\st L \mapsto V * \distAs{V}{\dist}$.

The \ref{rule:hoare-alloc-v} triple is valid in \thelogic,
but how could we reuse it in a context where the initial value stored at the allocated location is not a deterministic value $v$
but rather a \emph{random} value variable~$V$?
To allow for this, \thelogic\ makes a natural---\emph{but significant}---generalization:
our triples are in fact not over single expressions~$e\in\Expr$,
but rather over random expression variables~$E \from \Outcomes \to \Expr$.
This allows us to directly prove:
\begin{proofrule}
  \infer*[lab=hoare-alloc]{}{
      \hoare{\TRUE} {\Alloc{V}} {\Ret L. L \mapsto V}
    }
  \label{rule:hoare-alloc}
\end{proofrule}
where ${\Alloc{V}}$ is shorthand for $(\fun\outc.{\Alloc{V(\outc)}})$.
\ref{rule:hoare-alloc} subsumes \ref{rule:hoare-alloc-v} by choosing $V=\fun\wtv.v$,
and precisely describes the effect of the command without requiring
ownership of the distribution of~$V$; in the case where we do initially own
$\distAs{V}{\dist}$, by \ref{rule:frame} we can also get
$
\hoare{\distAs{V}{\dist}} {\Alloc{V}} {\Ret L. L \mapsto V * \distAs{V}{\dist}}.
$

Using our triples,
we are able to provide a very general version of Iris's ``bind'' rule---a functional version of the Hoare Logic rule of sequential composition:
\begin{proofrule}
  \infer*[lab=hoare-bind]{
    \hoare{P}{E}{\Ret W.P'(W)}
    \\
    \forall W.\hoare{P'(W)}{K[W]}{\Ret V.Q(V)}
  }{
    \hoare{P}{K[E]}{\Ret V.Q(V)}
  }
  \label{rule:hoare-bind}
\end{proofrule}
Here $K[\hole]$ is a random evaluation context variable, and $K[\rvexpr]$ is short for $\fun \outcome. K(\outcome)[\rvexpr(\outcome)]$
Using \ref{rule:hoare-bind} we can introduce random variables for intermediate results; for example, to handle $\Alloc{(\sample(q))}$ we can first bind on the flip
and then apply the rules for allocation as sketched before:
\[
  \infer{
    \hoare {\TRUE} {\sample(q)} {\Ret V. \distAs{V}{\smash{\Ber{q}}}}
    \\
    \hoare {\distAs{V}{\smash{\Ber{q}}}} {\Alloc{V}} {\Ret L. L \mapsto V * \distAs{V}{\smash{\Ber{q}}}}
  }{
    \hoare {\TRUE} {\Alloc{(\sample(q))}} {\Ret L. \exists V.~ L \mapsto V * \distAs{V}{\smash{\Ber{q}}}}
  }
\]

\subsection{Probabilistic Frame-Preserving Updates}
\label{sec:key:update}

We have already discussed the importance of the \ref{rule:condop-lift} rule for supporting outcome-local reasoning.
Proving soundness of this rule, however, presents a serious challenge.
For example, \pcOL\ supports \ref{rule:condop-lift}, but only by relying on the simplifying effect of conflating disjointness and independence, while
\Bluebell\ only supports the rule by disabling
further framing via a heavy side-condition.
Given our new model of probabilistic resources, it is not at all obvious how and why the \ref{rule:condop-lift} rule should hold for us.

In \thelogic, \ref{rule:condop-lift} is in fact a special case of the
\ref{rule:c-lift} rule:
\begin{proofrule}
  \infer*[lab=c-lift]{
      \forall x \in \supp(\dist) \st
        \hoare {P(x)} E {\Ret V. Q(x)(V)}
    }{
      \hoare {\cond{\dist}{x}{P(x)}} E {\Ret V. \cond{\dist}{x}.Q(x)(V)}
    }
  \label{rule:c-lift}
\end{proofrule}
We show that \ref{rule:c-lift} is valid in \thelogic\ 
by strengthening the notion of frame-preservation embedded in the meaning
of triples to a new notion we call \emph{probabilistic frame preservation}.
We achieve the result in a modular fashion, following the Iris methodology,
which prescribes triples and their rules to be derived
from the combination of more primitive modalities and their rules:
triples are described in terms of a \emph{weakest precondition}~(WP) modality
$\wpre{e}{\Phi}$ which, in turn, is built around the \emph{update} modality
$\upd P$.
Below, we explain these components of Iris, and how we modified them in \thelogic.

Intuitively, Iris's WP modality $\wpre{e}{\Phi}$
holds of a resource~$a$ if every execution of~$e$ from $a$
results in a return value~$v$ and resource~$b$ which satisfies the postcondition
$\Phi(v)$.
In addition, $\wpre{e}{\Phi}$ insists that every possible frame of~$a$
is preserved by~$e$.
Using this modality, a triple $\hoare{P}{e}{\Phi}$ can be encoded as
$P \entails \wpre{e}{\Phi}$.

A crucial aspect of WP is that it transforms the current
resource into a new resource that satisfies the postcondition,
while preserving all frames (thus validating \ref{rule:frame}).
This frame property is represented by the valid entailment
$
  P * \wpre{e}{\Phi} \entails \wpre{e}{\Ret v.P * \Phi(v)}.
$
In Iris, WP is itself a derived formula, which uses the so-called
\emph{update modality} $ \upd P $.
Roughly speaking, $\upd P$
holds on a resource~$a$ if $P$ holds on some resource~$b$
such that every frame of~$a$ is also a valid frame of~$b$
(\ie we can ``update'' $a$ to $b$ so that $P$ holds without ``breaking'' any frame of $a$).
The frame property of WP is thus a reflection of the analogous frame property of updates:
$ P * \upd Q \entails \upd (P * Q) $.

In \thelogic, we similarly derive triples from WP, and WP from an update modality.
Thus, first of all, in order for triples to validate \ref{rule:c-lift},
the WP needs to satisfy the \ref{rule:c-wp-swap} rule:
\begin{proofrule}
  \infer*[lab=c-wp-swap]{}{
    \cond{\dist}{x}{\wpre{E}{\Ret V.Q(x)(V)}}
    \entails
    \wpre{E}{\Ret V.\strut\smash{\cond{\dist}{x}{Q(x)(V)}}}
  }
  \label{rule:c-wp-swap}
\end{proofrule}
The rule says that WP and conditioning commute;
consider for example the case where $\dist = \Ber{q}$.
The rule states that,
if~$E$ produces a distribution satisfying $Q(b)$ (for~$b\in\set{0,1}$)
from the current distribution conditioned on $X=b$ for some (implicit)~$X$,
then $E$, run from the unconditional distribution, produces a distribution
that satisfies $Q(0) \condop[q] Q(1)$.

Going one level down, since WP is built from update,
the proof of \ref{rule:c-wp-swap} boils down to proving a similar rule for update:
\begin{proofrule}
  \infer*[lab=c-upd-swap]{}{
    \cond{\dist}{x}{\upd P(x)}
    \entails
    \upd\cond{\dist}{x}{P(x)}
  }
  \label{rule:c-upd-swap}
\end{proofrule}
Unfortunately, if we only require the update to be frame-preserving,
as is done in standard Iris,
\ref{rule:c-upd-swap} would not hold.
The reason is that there are frame-preserving updates which do not commute with conditioning.
In particular, conditioning on some random variable $X$ (in some~$\psp$) is only possible if~$X$ is measurable in $\psp$.
Consequently, all updates performed under branches of this conditioning must at least preserve the measurability of~$X$ or else the conditioning ceases to be meaningful---and yet some frame-preserving updates do not!

Consider for example updating some $\psp_0$ to a coarser probability space $\psp_1$ by removing the information about the probability of $X=0$, such that $X$ is no longer measurable.
``Forgetful'' updates like this are always frame-preserving,
because if every event that is measurable locally is independent of every
event that the frame can measure, then if we make fewer events measurable
locally, the frame would remain independent of the local resource.
The update, however, \emph{does not preserve conditioning}:
conditioning on~$X$ in some~$\psp$ is only possible if~$X$ is measurable in $\psp$, and so all updates performed in branches of this conditioning must at least preserve the measurability of~$X$ or else the conditioning ceases to be meaningful.
To see why, observe the case where we start from some $\psp$ where~$X$ is measurable,
and distributed as a fair coin.
If we condition on~$X$ we would have
${\psp = \onehalf\cdot\psp|_{X=0} + \onehalf\cdot\psp|_{X=1}}$.
But if in the first branch of the conditioning we update $\psp|_{X=0}$ to the trivial probability space $\Triv{}$,
then~$X$ would no longer be measurable in
${\psp' = \onehalf\cdot\Triv{} + \onehalf\cdot\psp|_{X=1}}$, causing the conditioning to become senseless.

This is already a problem without considering dynamic allocation:
for instance, in \pcOL\ triples only insist on frame preservation,
so the logic was forced to impose side conditions that exclude
updates like the one above in crucial rules for conditioning.

In \thelogic, we propose a new definition of the update modality,
which imposes stronger constraints on the allowed updates
such that both the frame property of updates and \ref{rule:c-upd-swap} are validated.
The idea is to require updates to also preserve the validity of any
possible conditioning they might be part of.
We call such updates \emph{probabilistic frame-preserving} (or PFP) and show that
\begin{enumerate}
  \item Any standard frame-preserving update on elements of~$\monoid$
        (the underlying standard resources)
        can be lifted to a PFP update
        on the random resource variable component of our model.
        As special cases, we get that mutation and dynamic allocation
        of heap cells are PFP.
  \item Taking a new sample from a distribution is a
        PFP update to the probability space component of the model.
\end{enumerate}

\subsection{Iris's ``Authoritative'' Resource Algebra in the Probabilistic Setting}
\label{sec:key:auth}

To fully derive WP within the logic on top of update,
standard Iris makes essential use of a construction called
the ``authoritative resource algebra''~$\Auth(\monoid)$.
In essence, $\Auth(\monoid)$ consists of two kinds of elements,
built on top of the elements of an underlying resource algebra~$\monoid$
(\eg heaps):
(1) the \emph{authority} $\authfull g$,
  which gives exclusive ownership of a resource $g\in\monoid$ representing
  the ``global'' view of the state,
and (2) the \emph{fragment} $\authfrag a$,
  which gives ownership of a resource~$a\in\monoid$ representing
  local ownership of a piece of the state.
Fragments compose: $ \authfrag a_1 \mop \authfrag a_2 = \authfrag (a_1 \mop_\monoid a_2) $.
Exclusivity of the authority is represented by the fact that
$ \authfull g_1 \mop \authfull g_2 $ is undefined, and
thus in any valid separation of resources there is at most one authority.
Crucially, $\authfull g \mop \authfrag a$ is only valid if
$a$ is a ``sub-resource'' of $g$
(a concept specified in the definition of $\monoid$).

In \thelogic\ we build our model with an authoritative construction
on top of the resource algebra we sketched in \cref{sec:key:model}.
Intuitively, fragments $\authfrag{(\psp,R)}$ model
the user-facing assertions like~$L \mapsto V$, which describe ownership of local resources.
The authority
$\authfull{(\psp,R)}$ is used in the encoding of WP to constrain the local fragments
to be sub-resources of the global resource $(\psp, R)$.

The model we defined using the authoritative construction validates
all the desired rules, except for one (first proposed in~\cite{bluebell})
that represents a key property of conditioning:
\begin{proofrule}
  \infer*[lab=c-frame-bb]{}{
    P * \cond{\dist}{v}{Q(v)}
    \entails
    \cond{\dist}{v}{(P * Q(v))}
  }
  \label{rule:c-frame-bluebell}
\end{proofrule}
This rule is put to work in virtually every conditional proof.
For example, to condition on~$X$ in $\distAs{Y}{\dist'} * \distAs{X}{\dist}$,
one first applies
$
  \distAs{X}{\dist} \entails \cond{\dist}{x}{\sure{X=x}}
$
to obtain
$
  \distAs{Y}{\dist'} * \cond{\dist}{x}{\sure{X=x}}
$,
and then \ref{rule:c-frame-bluebell} to push the independent resource $\distAs{Y}{\dist'}$ under the conditioning modality, obtaining $\cond{\dist}{x}{(\sure{X=x} * \distAs{Y}{\dist'})}$.
This makes $Y$ available to each branch of the conditioning when we subsequently apply the rule \ref{rule:c-lift}.

In \thelogic, however, \ref{rule:c-frame-bluebell} in its full generality does not hold!
To see why, suppose we start from
$ \ownM{\authfull{(\psp,R)}} * \distAs{X}{\Ber{\onehalf}} $;
then we know that the global distribution~$\psp$ will see~$X$
having a fair coin distribution.
If we condition on~$X$,
obtaining
$ \ownM{\authfull{(\psp,R)}} * \cond{\Ber{\onehalf}}{x}{\sure{X=x}} $,
an application of \ref{rule:c-frame-bluebell} would bring us to
$ \cond{\Ber{\onehalf}}{x}{\ownM{\authfull{(\psp,R)}} * \sure{X=x}} $,
which is inconsistent, since in every branch of the conditioning
we would be simultaneously claiming that $X$ is deterministic (by $\sure{X=x}$)
and distributed as a fair coin (by the information in $\psp$).

Fortunately, we can resolve this issue by reinterpreting the meaning of the
authority: it no longer represents the global resource in absolute terms,
but only \emph{relative to the current conditional branch in which it is asserted}.
Specifically, we define the authoritative algebra so that conditioning
validates two new variations of \ref{rule:c-frame-bluebell}:
\begin{proofrules}
  \infer*[lab=c-frame]{}{
      \mfble P * \cond{\dist}{v}{Q(v)}
      \entails
      \cond{\dist}{v}{(\mfble P * Q(v))}
    }
  \label{rule:c-frame}

  \infer*[lab=c-auth]{
    \forall v\st Q(v) \entails \sure{X=v}
  }{
    \ownM{\authfull g} * \cond{\dist}{v}{Q(v)}
    \entails
    \cond{\dist}{v}{(\ownM{\authfull (g|_{X=v})} * Q(v))}
  }
  \label{rule:c-auth}
\end{proofrules}
The rule \ref{rule:c-frame} uses a new \emph{frameable} modality~$\mfble{P}$ (inspired by Spies et al.~\cite{daenerys}) to
guard the resource to be brought inside a conditioning:
it is the identity on fragments
($ {\ownM{\authfrag a} \entailEq \mfble \ownM{\authfrag a}} $),
but cannot wrap the authority
($ {\mfble \ownM{\authfull g} \entails \FALSE} $).
The rule \ref{rule:c-auth} allows for the sound transfer of global
information across a conditioning,
by performing the relevant conditioning on the global state.
The premise of the rule serves to identify the random variable $X$
being conditioned on (which is needed in order to condition the global
resource $g$ on $X$).
The two rules together cover all the user-facing
uses of \ref{rule:c-frame-bluebell} (which would only involve fragments),
as well as the new manipulations on the authority required when proving properties of WP.

Equipped with this machinery, \thelogic\ supports the encoding of a rather standard-looking
definition of WP within the base probabilistic separation logic,
which can be manipulated abstractly using our new rules of conditioning and update.

\section{The \thelogic\ Model}
\label{sec:model}

This section describes the model of \thelogic. We start by recapping prior concepts (Preliminaries) and then describe the specifics of our new model.

\subsection{Preliminaries}
\label{sec:prelim}

\paragraph{Probability spaces.} In this paper, we use probability spaces for expressing partial information: A probability space does not give probability to every individual outcome, but instead to sets of outcomes (\emph{events}). The probability of an event $\event$ reflects the probability that the actual outcome lies within $\event$.

A \emph{probability space} $\psp$ is a tuple $\psp = (\Outcomes, \salg, \dist)$. The \emph{outcome space} $\Outcomes$ is the set of all probabilistic outcomes, the \emph{$\sigma$-algebra} $\salg \subseteq 2^\Outcomes$ is the set of all \emph{events}: subsets of $\Outcomes$ whose probabilities can be measured, and the \textit{distribution} $\dist: \salg \rightarrow [0,1]$ gives a probability to every event in $\salg$. When $\Outcomes$ is clear from the context, we often omit it.

Formally, the $\sigma$-algebra must contain $\Outcomes$ and be closed under countable unions, intersections and complements. The set of all $\sigma$-algebras over $\Outcomes$ is denoted $\SigAlg(\Outcomes)$.
The distribution $\dist: \salg \rightarrow [0, 1]$ must satisfy $\dist(\Outcomes) = 1$ and is required to be \emph{countably additive}, meaning that the probability of a countable union of pairwise disjoint events is the sum of the events' probabilities.

The \emph{support} of $\dist$ is the set of outcomes not deemed impossible. It is the smallest event $\event_{supp}$ such that $\dist(\event_{supp}) = 1$, if such an event exists (which it does for the probability spaces we work with).

The probability space $\psp = (\salg, \dist)$ is \textit{complete} if for all events $\event \in \salg$ with $\dist(\event) = 0$ and all $\event' \subseteq \event$, it holds that $\event'\in\salg$. Completeness reflects the intuition that if an event has probability 0, we should know that all sub-events also have probability 0.

Given $\psp_1=(\salg_1, \dist_1),\psp_2=(\salg_2, \dist_2)$
with $\salg_1 \in \SigAlg(\Outcomes_1)$
and  $\salg_2 \in \SigAlg(\Outcomes_2)$,
their product~$ \psp_1 \pprod \psp_2 $ is the probability space
over $\Outcomes_1 \times \Outcomes_2$ with
\salgebra{} $\sigcl{\set{\event_1 \times \event_2 | \event_1 \in \salg_1, \event_2 \in \salg_2}}$,
and distribution $(\dist_1 \pprod \dist_2)(\event_1 \times \event_2) = \dist_1(\event_1)\cdot\dist_2(\event_2)$.

For $\salg_1 \in\SigAlg(\Outcomes)$, a restriction of $\dist$ to $\salg_2\in\SigAlg(\Outcomes)$ ($\salg_2 \subseteq \salg_1$) drops the information about all events not in $\salg_2$: it is the distribution $\dist|_{\salg_2}$ with $\dist|_{\salg_2}(\event) = \dist(\event)$ for all events in $\salg_2$.
The \emph{extension order} relates $(\salg_1, \dist_1) \sqsubseteq (\salg_2, \dist_2)$ iff $\salg_1 \subseteq \salg_2$ and $\dist_1 = \dist_2|_{\salg_1}$: intuitively, $(\salg_2, \dist_2)$ contains all information that $(\salg_1, \dist_1)$ also contains and possibly more.

\paragraph{Random variables.} Given a set of outcomes, $\Outcomes$, a $T$-valued random variable is a function $X: \Outcomes \to T$. The random variable is \textit{measurable} in a probability space $(\salg, \dist)$ if we can give probability to every possible value of $X$, \ie, if $\forall v \in T.~X^{-1}(v) \in \salg$. If $X$ is measurable, we say that $X$ is \textit{distributed as} $\dist_X$ if $\forall v \in \supp(\dist_X).~\dist(X^{-1}(v)) = \dist_X(v)$.

\paragraph{Independent product.} Intuitively, the independent product combines the information of two independent probability spaces over the same outcome space into one. Given $\psp_1 = (\salg_1, \dist_1)$ and $\psp_2 = (\salg_2, \dist_2)$, let $\sigma(\salg_1\cup\salg_2)$ be the smallest $\sigma$-algebra containing all elements of $\salg_1 \cup \salg_2$.  Then, we say that $\psp_1 \iprod \psp_2 = (\sigma(\salg_1\cup\salg_2), \dist)$ is the \emph{independent product} of $\psp_1$ and $\psp_2$ if $\dist(\event_1 \cap \event_2) = \dist_1(\event_1) \cdot \dist_2(\event_2)$ for any $\event_1 \in \salg_1$ and $\event_2 \in \salg_2$. The independent product does not exist when there are $\event_1, \event_1' \in \salg_1$ and $\event_2, \event_2' \in \salg_2$ such that $\event_1 \cap \event_2 = \event_1' \cap \event_2'$ but $\dist_1(\event_1) \cdot \dist_2(\event_2) \neq \dist_1(\event_1') \cdot \dist_2(\event_2')$. The independent product is unique if it exists \cite{lilac}. 

\subsection{Our Resource Algebra}
We formulate our model as an \textit{ordered resource algebra} (ORA) \cite{mosel}, a structure that generalizes Kripke resource monoids \cite{kripkemonoid} and is used in many Iris developments. A model that is an ORA immediately justifies the soundness of all standard separation logic rules. An ORA consists of:
a \emph{carrier type} $\monoid$;
the \textit{resource composition} $\cdot : \monoid \to \monoid \to \monoid$;
a \emph{validity predicate} $\mval: \monoid \to \Prop$, which defines which elements of the carrier type are valid and is used to model partiality of the composition;
a \textit{resource order} $\mincl$, which is a preorder over $\monoid$ and denotes that resources contains `more information' than another;
a \textit{persistent core} $\mcore{\!-\!} : \monoid \to \monoid$ that captures the duplicable part of a resource;
and a \textit{unit} $\munit$ which is the neutral element of $\cdot$.

We build our model in two layers: First we define an ORA on probability spaces, and then use it to define the ORA of our full model. 

\paragraph{Our ORA on probability spaces}
The core of our model is an ORA defined on probability spaces over $\Outcomes = \Rid \to \Bool$ for a fixed, countable set of randomness
IDs, $\Rid$.  Our probability spaces can always be factored into a finite and trivial part: they have the form ~$\psp = \psp_R
\pprod \Triv{(\Rid\setminus R) \to \Bool}$, where~$R
\subs_{\mathsf{fin}} \Rid$, $\psp_R$ is a probability space over the
finite set $R \to \Bool$, and $\Triv{\Outcomes} \in
\ProbSp(\Outcomes)$ is the \salgebra\ $ \set{\Outcomes,\emptyset} $
equipped with the trivial distribution.

Our Rocq implementation relies on a concrete representation of such
spaces that keeps track of~$R$ and $\psp_R$ explicitly. 
To simplify notation, here, we omit the details of this representation and present definitions directly on the full outcome space $\Rid \to \Bool$.
The details of our representation can be found in \appendixref{sec:full-model}{B}.

Our ORA on probability spaces is a reformulation of Lilac's and Bluebell's Kripke resource monoids.
Crucially, the ORA's resource composition is the independent product of probability spaces, and its persistent core is the \emph{sure} subspace (events with probability $0$ or $1$) of the probability space.
The sure subspace of $\psp$ is always independent of $\psp$, and can thus be duplicated arbitrarily.
\begin{align*} 
  &\PSpRA &&\eqdef \{\psp \mid \psp \text{ is a complete probability space over } \Rid \to \Bool \}  \cup \{ \mundef \}\\
  &\mval(a) &&\eqdef a \neq \mundef \\
  &\psp_1 \cdot \psp_2 &&\eqdef \psp_1 \iprod \psp_2 \qquad \text{if it exists, else } \mundef\\
  &\mcore{(\salg, \dist)} &&\eqdef (\{\event \in \salg \mid \dist(\event) = 0 \lor \dist(\event) = 1\}, \dist) \\
  &\munit &&\eqdef \Triv{\Rid \to \Bool} \\
  &\psp \mincl \psp' &&\eqdef \psp \extorder \psp' \lor \psp' = \mundef
\end{align*}

  \paragraph{Random resource variables and authoritative component}
  \colorlet{authcolor}{black}
  \colorlet{rvcolor}{black}
  Next, we present our full model, the ORA $\PSpAuth{\monoid}$.
Given an Iris resource algebra $\monoid$, an element of $\PSpAuth{\monoid}$ is a triple of a probability space over $\Outcomes$,
  an $\monoid$-valued random variable (called a \emph{random resource variable}) of type $\Outcomes \to \monoid$, and an optional authoritative component.
  The random resource variable component lifts all resource algebra connectives of $\monoid$ pointwise to $\Outcomes \to \monoid$.
  The authoritative component behaves like Iris' authoritative resource algebras: It is an upper bound on the probability space (in the extension order $\mincl$), and it is \textit{exclusive} --- two different authorities cannot be composed. 
Following Iris terminology, the probability space component is also called a \textit{fragment} of the authoritative part.
{\allowdisplaybreaks
\begin{align*}
      \PSpAuth{\monoid} &\eqdef \PSpRA  \times {\color{rvcolor}\rv{\monoid}} \times \color{authcolor}{\moption(\PSpRA)}\\
       (\fraginstance, {\color{rvcolor}{\rvRes}}, {\color{authcolor}\authinstance}) \equiv (\fraginstance', {\color{rvcolor}\rvRes'}, {\color{authcolor}\authinstance'}) &\eqdef \fraginstance = \fraginstance' \land {\color{authcolor}\authinstance=\authinstance'} \land {\color{rvcolor}\forall \outcome \in \supp(\psp).~ \rvRes(\outcome) \equiv \rvRes'(\outcome)}\\
      (\fraginstance, {\color{rvcolor}{\rvRes}}, {\color{authcolor}\authinstance}) \cdot (\fraginstance', {\color{rvcolor}\rvRes'}, {\color{authcolor}\authinstance'}) &\eqdef (\fraginstance \cdot \fraginstance', \color{rvcolor}{\fun \outcome.~\rvRes(\outcome) \cdot \rvRes'(\outcome)}, \color{authcolor}{\authinstance \authop \authinstance'}) \\
      \mval(\fraginstance, {\color{rvcolor}{\rvRes}}, \None) &\eqdef \color{rvcolor}\exists \fraginstance'.~\mval(\fraginstance') \land \fraginstance \mincl \fraginstance' \land \forall \outcome \in \supp(\fraginstance').~\mval(\rvRes(\outcome)) \\
      \mval(\fraginstance, {\color{rvcolor}{\rvRes}}, {\color{authcolor}\psp_a}) &\eqdef {\color{authcolor}\mval(\psp_a) \land \fraginstance \mincl \psp_a} \land \color{rvcolor}{\forall \outcome \in \supp(\psp_a).~ \mval (\rvRes (\outcome))} \\
\mcore{(\fraginstance, {\color{rvcolor}\rvRes}, {\color{authcolor}\authinstance})} &\eqdef (\mcore \fraginstance, {\color{rvcolor}\fun \outcome.~ \mcore{\rvRes(\outcome)}}, \color{authcolor}{\None}) \\
      \munit &\eqdef (\munit, {\color{rvcolor}\fun. \munit}, {\color{authcolor}{\None}}) \\
      (\fraginstance, {\color{rvcolor}\rvRes}, {\color{authcolor}{\authinstance}}) \mincl (\fraginstance', {\color{rvcolor}\rvRes'}, {\color{authcolor}{\authinstance'}}) &\eqdef \fraginstance \mincl \fraginstance' \land {\color{rvcolor}\forall \outcome \in \supp(\fraginstance').~ \rvRes(\outcome) \mincl \rvRes'(\outcome)}
      \land 
      \color{authcolor}{(\authinstance \neq \None \implies \authinstance = \authinstance')} 
    \end{align*}   }
    where $\color{authcolor}\Some a \authop \None = \Some a =  \None \authop \Some a$, $\color{authcolor}\Some \psp_a \authop \Some \psp_b = \mundef$, and $\rv{\monoid} := (\Rid \to \Bool) \to \monoid$.

We declare two resources equivalent if their random resource variable
components are almost surely equivalent, i.e. equivalent on all
outcomes that are not probability 0.

Our definition of validity for $(\fraginstance,
{\color{rvcolor}{\rvRes}}, \None)$ may be a bit surprising: One might
expect a simpler formulation like $\mval(\psp) \land \forall \outcome
\in \supp(\psp).~\mval(R(\outcome))$. However, note that in an ORA,
validity must be \textit{downwards-closed}, \ie, an extension of an
invalid element must be invalid. Our definition forces this closure,
while the simpler definition does not. To see this, note that going up
in the order $\mincl$ may shrink the support of the probability
space. For example, the support of the trivial probability space
$\Triv{\Rid \to \Bool}$ is all of $\Outcomes$, but in an extension
where $\iota = 1$ with probability $1$, the support does not include
outcomes with $\iota=0$. Hence, if the original ORA element was
invalid only because the random resource variable mapped $\iota = 0$
to an invalid RA element, then the extension will be valid, violating
downward closure.

For validity with an authority, our simple definition is sufficient
since the authority remains unchanged when going up in the order.

We define projections of our ORA: $\projout{\projfrag}{(\fraginstance, \rvRes, \authinstance)} = \fraginstance, \projout{RV}{(\fraginstance, \rvRes, \authinstance)} = \rvRes, \projout{\projauth}{(\fraginstance, \rvRes, \authinstance)} = \authinstance$.

\subsection{Conditioning and Weighted Sum}
\label{sec:cond-wsum}

In this section, we provide two equivalent definitions of the
conditioning modality. The first is a top-down definition that
directly captures the intuition of decomposing a probability space
along a random variable. The second is a bottom-up characterization in
terms of a weighted sum operation that we use later to define
probabilistic frame-preserving updates and to prove that updates can
be lifted out of conditioning (rule \ref{rule:c-wp-swap}, Section
\ref{sec:key:update}).

\paragraph{Definition of the conditioning modality.} For a probability space $\psp = (\salg, \dist)$ and an event $\event \in \salg$ with $\dist(\event) > 0$, the \emph{conditioning} $\psp|_\event$ is defined as $(\salg, \dist')$, where $\dist'(\eventB) = \dist(\eventB \cap \event) / \dist(\event)$ for all $\eventB \in\salg$. Intuitively, $\psp|_\event$ rules out outcomes outside $\event$ and renormalizes the remaining probabilities. Building on this notion, we define the conditioning modality (here, $\condpure{\None}{\event} \eqdef \None$ for all $\event$):
\begin{align*}
(\cond{\dist}{\wsumInx}{\pred(\wsumInx)})(\fraginstance,\rvRes,\authinstance) &\eqdef \exists X.~ X \text{ is distributed as } \dist \text{ in } \fraginstance \land 
\forall \wsumInx \in \supp \dist.~\pred(\wsumInx)(\condpure{\fraginstance}{X = \wsumInx},~\rvRes, \condpure{\authinstance}{X= \wsumInx})
\end{align*}
where $\dist$ is a finite support distribution that assigns probability to every outcome, \ie a map from outcomes to $\mathbb{R}$.

In the above definition, both the fragment and the authority are decomposed along a random variable $X$. For the fragment, this is standard and analogous to Lilac's or Bluebell's conditioning. Regarding the authority, recall the discussion from Section \ref{sec:key:auth}: while in traditional Iris, the authority is interpreted as a ``global ground truth'' and never decomposed, Amaryllis's authority is relative to the current branch $\wsumInx$. It represents the ground truth probability space for branch $v$, which is the global probability space conditioned on $X=\wsumInx$. The random resource variable does not need modification: the cases that do not fall in the current branch automatically become irrelevant since they are probability 0 in $\psp|_{X=\wsumInx}$.

\paragraph{Alternative characterization.} In order to prove \ref{rule:c-upd-swap}, we need a way to combine resources that were updated locally under conditioning into one globally updated resource. For this, we will proceed to define a weighted sum operation $\wsumfn : \forall A.~\Dist(A) \to (A \to \PSpAuth{\monoid}) \to \PSpAuth{\monoid}$ that combines resources of multiple probabilistic branches into a single resource. The weighted sum will also yield an alternative characterization of conditioning:
$$(\cond{\dist}{\wsumInx}{\pred(\wsumInx)})(x) \iff \exists \kernel.~ \wsum{\dist}{\kernel} \mincl x \land \forall \wsumInx \in \supp \dist.~\pred(\wsumInx)(\kernel(\wsumInx)) $$

To ensure the equivalence between both definitions, the weighted sum enforces the existence of a random variable separating the branches. The individual branches $\kernel(v)$ then relate to the branches of the top-down definition in the following way: We have shown that, for every valid $\wsum{\dist}{\kernel}$, there exist $X$ and $R$ such that $\forall \wsumInx \in \supp \dist.~ \kernel(\wsumInx) \equiv ((\projout{\projfrag}{\wsum{\dist}{\kernel}})|_{X=\wsumInx}, ~\rvRes,~ \projout{\projauth}{(\wsum{\dist}{\kernel})|_{X=\wsumInx}})$.

We will ultimately (in Section~\ref{sec:pfpu}) require all resource updates to not only preserve frames but also \textit{preserve conditionings} by requiring them to preserve the validity of all weighted sums: if the validity is preserved, an update done locally can always be lifted to a global update by applying the weighted sum operation.

In the following, we will define $\wsumfn$ first on probability spaces, and then lift it to our full ORA.

\paragraph{Weighted sum on probability spaces}
We define the weighted sum on probability spaces as follows.
\[
\wsumPSp{\dist}{\kernel} \eqdef
\begin{cases}
  \Inters_{\wsumInx \in \supp(\dist)} \projout{\salg}{(\kernel(\wsumInx))},~\fun \event.~\sum_{\wsumInx \in \supp(\dist)} \dist(\wsumInx) \cdot \projout{\dist}{(\kernel(\wsumInx))}(\event)  & \text{if } \hasrv(\dist,\kernel), \\
  \mundef & \text{otherwise.}
\end{cases}
\]\begin{align*}
  	\hasrv(\dist, \kernel) &:= \forall \wsumInx, w \in \supp{\dist}.~\wsumInx \neq w \implies \supp(\kernel(\wsumInx)) \cap \supp(\kernel(w)) = \emptyset 
\end{align*}
The weighted sum is defined if the supports of $\kernel$'s branches are disjoint, as codified in the condition $\hasrv(\dist, \kernel)$. This is equivalent to requiring that $\mu$ be the distribution of \emph{some} random variable, and that $\kernel$ be the branches defined by the possible values of that random variable. If this is the case, we define the $\sigma$-algebra of the weighted sum to be the set of events that are measurable in all branches and apply the standard $\bind$ operation of the Giry monad for the probabilities.

\paragraph{Weighted sum on $\PSpAuth{\monoid}$}
We extend the definition of weighted sum to our ORA. To take the weighted sum of random resource variables, consider a specific outcome $\outcome \in \Outcomes$. Since the supports of $\kernel$'s branches must be pairwise disjoint (see above), there is at most one branch $\wsumInx$ such that $\outcome$ has nonzero probability in that branch. If such a $\wsumInx$ exists, define $\rvRes(\outcome) = \projout{RV}{(\kernel(\wsumInx))}(\outcome)$.

Formally, define $\getrv(\dist, \kernel)(\outcome)$ as the unique $\wsumInx \in \supp(\dist)$ such that $\outcome \in \supp \projout{\projfrag}(\kernel(\wsumInx))$ if such a $\wsumInx$ exists, else let it be arbitrary. Then,
\begin{align*}
  &\wsum{\dist}{\kernel} \eqdef (\fraginstance, \rvRes, \authinstance) \quad
  &\fraginstance \eqdef \wsumPSp{\dist}{\fun \wsumInx. \projout{\projfrag}{(\kernel~\wsumInx)}} 
  \quad \rvRes \eqdef  \fun \outcome. \projout{RV}{(\kernel(\getrv(\dist,\kernel)(\outcome)))}~ \outcome
\end{align*}
Note that the random resource variable $\rvRes_\wsumInx$ of any branch $\wsumInx$ is almost surely equal to $\rvRes$ in the probability space of branch $\wsumInx$ (in other words, $\rvRes$ and $\rvRes_\wsumInx$ only differ on outcomes that are probability 0 in branch $\wsumInx$). This is why the top-down definition of conditioning, in which all branches have the same $\rvRes$, is equivalent to the bottom-up characterization based on the weighted sum.

We require the authority to either exist in all branches of $\kernel$ or none of them: restricting the possible decompositions to those cases is enough for compatibility with the top-down definition.
If it exists in all of them, we define the resulting authority as the $\wsumfn_\PSpRA$ of the individual authorities. 
\begin{align*}
  &\authinstance \eqdef \None &&\text{ if } \forall \wsumInx.~\mval(\kernel(\wsumInx)) \land \projout{\projauth}{(\kernel(\wsumInx))} = \None \\
  &\authinstance \eqdef \Some \wsumPSp{\dist}{\kernel'} &&\text{ if } \forall \wsumInx.~\mval(\kernel(\wsumInx)) \land \projout{\projauth}{(\kernel(\wsumInx))} = \kernel' (\wsumInx) \\
  &\authinstance \eqdef \mundef &&\text{else}
\end{align*}

We have shown that with this definition of the weighted sum, the two characterizations of conditioning above are equivalent.

\subsection{Probabilistic Frame-Preserving Updates} \label{sec:pfpu}
Next, we define PFP updates, which generalize Iris updates. Like Iris updates, PFP updates preserve frames.
Additionally, we require PFP updates to \emph{preserve conditioning}, \ie the validity of weighted sums.
Formally, a PFP update from resources $x$ to $x'$ preserves conditioning if
$$\forall \dist, \kernel, \wsumInx.~\mval(\wsum{\dist}{\kernel[\wsumInx \mapsto x]}) \implies \mval(\wsum{\dist}{\kernel[\wsumInx \mapsto x']})$$

We can now define \emph{probabilistic contexts}, $\pctx{M}$, comprising the $\wsumfn$ operator and the resource composition ($\cdot$), and define context application, $\pctxFill{\pctx{M}}{x}$.
\begin{align*}
  \pctx{\monoid} &\eqdef \pctxHole \mid \pctxFrame{cx}{\pctx{\monoid}} \mid \pctxCond{\dist}{\kernel}{\wsumInx}{\pctx{\monoid}} 
\end{align*}
\vspace{-0.7cm}\begin{align*}
  \pctxFill{\pctxHole}{x} \eqdef x \quad
  \pctxFill{\pctxFrame{cx}{\pctx{\monoid}}}{x} \eqdef cx \cdot \pctxFill{\pctx{\monoid}}{x} \quad
  \pctxFill{\pctxCond{\dist}{\kernel}{\wsumInx}{\pctx{\monoid}}}{x} \eqdef \wsum{\dist}{\kernel[\wsumInx \mapsto \pctxFill{\pctx{\monoid}}{x}]}
\end{align*}

Equipped with these definitions, we define \emph{probabilistic frame-preserving (PFP) updates}, as those resource updates that preserve validity in all contexts:
$$\pvs{} P \eqdef \fun a.~\forall \pctx{\monoid}.~\mval{(\pctxFill{\pctx{\monoid}}{a})} \implies 
  \exists a'.~\mval(\pctxFill{\pctx{\monoid}}{a'}) \land P(a') $$
This definition of updates now validates both \ref{rule:frame} due to its frame-preservation and \ref{rule:c-upd-swap} due to its preservation of weighted sums.
Moreover, we have shown that allocating new randomness IDs is a PFP update, which is a consequence of the following more general theorem:

\begin{theorem}[Adding information is a PFP update] \label{thm:upd-ext}
  \begin{align*}
    \forall \pctx{M}.~ \mval(\psp_a \cdot \psp) \rightarrow \mval(\pctxFill{\pctx{M}}{(\munit, \munit, \psp_a)}) \rightarrow
    \mval(\pctxFill{\pctx{M}}{(\psp, \munit, \psp_a \cdot \psp)})
  \end{align*}
\end{theorem}

In addition, we have shown that frame-preserving updates on the random resource variable (including non-deterministic updates) can be lifted to PFP updates:

\begin{theorem}[Lifting frame-preserving updates] \label{thm:upd-lift}
\begin{align*} 
  &\left(\forall \outcome~ cx.~ \mval{(\rvRes(\outcome)\cdot cx)} \implies \exists x'.~\mval(x' \cdot cx) \land \Phi(\outcome)(x')\right) \implies \\
  &\forall \pctx{M}.~\mval(\pctxFill{\pctx{M}}{(\munit, \rvRes, \None)}) \implies \exists \rvRes'.~\mval(\pctxFill{\pctx{M}}{(\munit, \rvRes', \None)}) \land \forall \outcome.~\Phi(\outcome)(\rvRes'(\outcome))
\end{align*}
\end{theorem}
This theorem says that if a random resource variable $\rvRes$ can be updated in the Iris frame-preserving sense to a resource that satisfies the predicate $\Phi$ in all probabilistic branches, then the probabilistic resource $(\munit, \rvRes, \None)$ can be PFP updated (in the sense defined above) to satisfy $\Phi$ in all branches.
Intuitively, the theorem holds in our model because all operations on random resource variables are defined per branch (per outcome). A weighted sum has no effect on non-zero probability branches of the random resource variable, and ignores zero probability branches. Thus, an update in a branch of the random resource variable, whether probability 0 or not, will always preserve the validity of arbitrary weighted sums.
See our Rocq development~\cite{artifact} for the formal proof.

 \section{Language and Semantics}

Like Iris, Amaryllis is parametric in the language used. It can be instantiated with any language that admits a single-step semantics of the following form:
\[
  \pstep~ \subseteq (\Expr \times \State) \times (\Expr \times \State \times \Expr \times \State \times \mathbb{R})
\]
where $\expr, \pstate \pstep \expr_1, \pstate_1, \expr_2, \pstate_2, p$ (for $0 < p \leq 1$) can be read as: $\expr, \pstate$ steps to $\expr_1, \pstate_1$ with probability $p$, and to $\expr_2, \pstate_2$ with probability $1-p$.

For concreteness, this paper considers the language \lang, a variant of Iris's HeapLang language, extended with real numbers~$r$ and a sampling primitive $\sample(e)$ that samples a boolean value from the Bernoulli distribution with bias $e$. \lang~ supports higher-order functions and higher-order state.
\begin{align*}
  \Val \ni \val \bnfdef&~ z \in \Int \mid r \in \Real \mid b \in \Bool \mid \loc \in \Loc \mid \Rec{f}{x} = \expr \mid (\val_1, \val_2) \\
  \Expr \ni \expr \bnfdef&~ \val \mid \lvar{x} \mid \expr_1 ~ \expr_2 \mid \Alloc{\expr} \mid \Get{\expr} \mid \Set{\expr_1}{\expr_2} \mid \expr_1 + \expr_2 \mid \expr_1 \leq \expr_2 \mid \dots \\
  &\mid(\expr_1, \expr_2)\mid \Fst{\expr} \mid \Snd{\expr}\mid \If \expr_1 then \expr_2 \Else \expr_3 \mid \sample~ \expr
\end{align*}
We use notation $\fun \lvar{x}. e \eqdef \Rec{\_}{x} = e$, and $\Let {\lvar{x}} = e_1 in e_2 \eqdef (\fun \lvar{x}. e_2) e_1$.

For instance, we have that
$\sample(p), \pstate \pstep \True, \pstate, \False, \pstate, p$.
All standard non-probabilistic steps $\expr, \pstate \pstep_{np} \expr', \pstate'$ are lifted to probabilistic steps $\expr, \pstate \pstep \expr', \pstate', \expr_2, \pstate_2, 1$ for all $\expr_2, \pstate_2$. For example, $\Alloc v, \pstate \pstep l, \pstate[l \mapsto v], \expr_2, \pstate_2, 1$ for all $l \notin \dom \pstate, \expr_2, \pstate_2$.
\lang's single-step semantics is defined as a \textit{contextual} language in the standard way \ie for any evaluation context $\lctx \in \Lctx$ it holds that $\expr, \pstate \pstep \expr_1, \pstate_1, \expr_2, \pstate_2, p$ implies that $\lctx[\expr], \pstate \pstep \lctx[\expr_1], \pstate_1, \lctx[\expr_2], \pstate_2, p$.

This single-step semantics can be lifted to a big-step semantics $\stdstep~ \subseteq (\Expr \times \State) \times \mathbb{D}(\Val \times \State)$ in a standard way ~\cite{probsem1, probsem2, probsem3}:
\begin{proofrules} 
  \infer*{}{\val, \sigma \stdstep \dirac{\val, \sigma}}
  
  \infer*{\expr, \sigma \pstep \expr_1, \sigma_1, \_, \_, 1\quad \expr_1, \sigma_1 \stdstep \dist}{\expr, \sigma\stdstep\dist}

  \infer*{e, \sigma \pstep e_1, \sigma_1, e_2, \sigma_2, p\quad e_1, \sigma_1\stdstep \mu_1\quad e_2, \sigma_2\stdstep \mu_2}{e, \sigma\stdstep \bind(\Ber{p}, (\lambda n, \text{if } n = 1\text{ then } \mu_1\text{ else }\mu_2))}
\end{proofrules}
where the \emph{Dirac distribution} $\dirac{v}$ for $v \in \Outcomes$ is defined as $\dirac{v}(v) = 1$ and $\dirac{v}(w) = 0$ for all $w \neq v$.

This is the semantics we will use in our adequacy theorem. However, we still need to connect the operational semantics to our model and weakest precondition: the model is based on indexed valuations, and the WP, as discussed in \ref{sec:key:specs}, contains a random expression variable in the middle of the triple. Therefore, in the following, we also lift the single-step semantics to a collecting semantics on indexed valuations that matches this structure:
$$\bigstepfn~ \subseteq (\psp \times \rv{\Expr \times \State}) \times (\psp \times \rv{\Val \times \State})$$

The standard semantics' big-step judgments $\expr, \sigma \stdstep \dist$ can be shown equivalent to the judgment $\bigstep{\emptyset}{\fun . (\expr, \sigma)}{\psp}{\rvcfg}$ by expressing $\dist$ as the \textit{push-forward} of $\psp$ along $\rvcfg$. Formally:

\begin{theorem}[Equivalence of Big-Step Semantics] Both of the following hold:
  \begin{itemize}
  \item $\expr, \sigma \stdstep \dist \rightarrow \exists \psp, \rvcfg.~ \dist = (\projout{\dist}{\psp}) \circ \rvcfg^{-1} \land \bigstep{\emptyset}{\fun . (\expr, \sigma)}{\psp}{\rvcfg}$
  \item $\bigstep{\emptyset}{\fun . (\expr, \sigma)}{\psp}{\rvcfg} \rightarrow \expr, \sigma \stdstep (\projout{\dist}{\psp}) \circ \rvcfg^{-1}$
\end{itemize}
\end{theorem}

\subsection{Collecting Semantics on Indexed Valuations}
Intuitively, $\bigstepfn$ is defined as follows: For each step $\pstep$ that is executed in some branch, a new randomness ID is allocated and the random configuration variables are updated accordingly. For example, consider the program $\Let x = \flip in \Alloc{(\flip + x)}$, executed in an initially empty state and empty probability space. The first flip will create an RID $\rid_1$ and leave us with the random configuration variable $\fun \outcome. (\Alloc{(\flip + \outcome(\rid_1))}, \emptyset)$ \footnote{To simplify the presentation, we ignore the trivial coin flips caused by deterministic steps here.}. The second flip creates two new randomness IDs, $\rid_2$ and $\rid_3$, one for each of the two outcomes of the first flip, leaving us with the random configuration variable $\fun \outcome. ((\text{if } \outcome(\rid_1) \text{ then } \outcome(\rid_2) \text{ else } \outcome(\rid_3)) + \outcome(\rid_1), \emptyset)$.
The last step, which is the allocation of a reference, will result in the random variable $\fun \outcome. (L(\outcome),~ L(\outcome) \mapsto (\text{if } \outcome(\rid_1) \text{ then } \outcome(\rid_2) \text{ else } \outcome(\rid_3)) + \outcome(\rid_1))$. Note that this is not the only possible reduction as the RIDs and the RV values for probability 0 branches are chosen nondeterministically.

 Formally, $\bigstepfn$ is defined in terms of the \textit{atoms} of the original probability space:  An event of a probability space is called an atom if it has nonzero probability and all strict sub-events have probability 0. We write $\atoms{(\psp)}$ for the set of all atoms of $\psp$. With this, we now define an auxiliary relation $\mstep$ by formally lifting the single-step relation to indexed valuations.
\begin{align}
& \psp_1, \rvcfg_1 \mstep \psp_2, \rvcfg_2  :=
\exists X,\rid,e_1,\sigma_1,e_2,\sigma_2,p. \label{eq:existentials}\\
&\quad X \in \atoms(\psp_1) ~\land \label{eq:atom} \\
&\quad (\exists \psp_1'.~ \psp_1 = \psp_1'  \pprod \Triv{\{\rid\} \to \Bool}) \label{eq:free-rid}~\land\\
&\quad \psp_2 = \wsum{(\projout{\mu}{\psp_1})|_{\atoms(\psp_1)}}{
 \lambda Y.~\psp_1|_Y\cdot \mathsf{if}~X = Y~\mathsf{then}~ \distAs{\rid}{\Ber{p}}~\mathsf{else}~\varepsilon}~\land \label{eq:alloc-rid}\\
&\quad (\forall \outcome \in X.~ \rvcfg_1(\outcome) \leadsto e_1, \sigma_1, e_2, \sigma_2, p)~\land \label{eq:correct-step}\\
&\quad (\forall \outcome \in X.~ \rvcfg_2(\outcome[\rid \mapsto \True]) = (e_1, \sigma_1) ~\land (p \neq 1 \rightarrow \rvcfg_2(\outcome[\rid \mapsto \False]) = (e_2, \sigma_2))) ~\land \label{eq:upd-rv}\\
&\quad \forall Y \in \atoms(\psp_1), \rho \in Y.  Y \neq X \rightarrow \rvcfg_1(\rho) = \rvcfg_2(\rho) \label{eq:no-change}
\end{align}

Every atom of $\psp_1$ corresponds to a concrete possible outcome of the coin flips that happened so far. In the definition of $\mstep$, the atom $X$ that will step next and the RID $\rid$ to be allocated are both chosen nondeterministically (Lines \ref{eq:existentials}, \ref{eq:atom}). Line \ref{eq:free-rid} ensures that $\rid$ has not been allocated yet, i.e. $\psp_1$ contains no information on $\rid$. Line \ref{eq:alloc-rid} specifies how to update the probability space with information on the new $\rid$: the atom $X$ is split into two, one for the case that $\rid$ is true, and one for the case that $\rid$ is false. Formally, this is achieved by replacing the $X$ branch of $\psp_1$ (\ie $\psp_1|_X$) with the independent product of $\psp_1|_X$ and the probability space stating the distribution of $\rid$ (here denoted as $\distAs{\rid}{\Ber{p}}$, under a slight abuse of notation), while leaving the other branches unchanged. Note that the probability space specifying $\distAs{\rid}{\Ber{p}}$ is guaranteed to be independent of $\psp_1|_X$ by Line \ref{eq:free-rid}. Line \ref{eq:correct-step} ensures that for all outcomes in the current atom, the pair of expression and state (configuration) can indeed take the correct step.  Line \ref{eq:upd-rv} specifies how to update the random variable $\rvcfg_1$ to $\rvcfg_2$: If $\rid$ is true, the expression and state are now $\expr_1, \sigma_1$, and if $\rid$ is false (only possible if $p \neq 1$), expression and state are now $\expr_2, \sigma_2$. Lastly, Line~$\ref{eq:no-change}$ specifies that the remaining branches of the random variable remain unchanged. 

Using the relation $\mstep$, we can now build our big-step semantics, $\bigstepfn$:
\begin{proofrules}
  \infer*[lab=big-step-refl] {
    \sure{\rvcfg_1 = \rvcfg_2}_\psp\\
    \forall \outcome.~\rvcfg_2(\outcome) \in \Val 
  } {
    \bigstep{\psp}{\rvcfg_1}{\psp}{\rvcfg_2}
  }

  \infer*[lab=big-step-step] {
    \psp_1, \rvcfg_1 \mstep \psp_2, \rvcfg_2\\
    \bigstep{\psp_2}{\rvcfg_2}{\psp_3}{\rvcfg_3}
  } {
    \bigstep{\psp_1}{\rvcfg_1}{\psp_3}{\rvcfg_3}
  }
\end{proofrules}
where $\sure{\rvcfg_1 = \rvcfg_2}_\psp$ is an almost-sure equality: $\forall \outcome \in \supp(\psp).~\rvcfg_1(\outcome) = \rvcfg_2(\outcome)$.

 \section{The \thelogic\ Logic}
\label{sec:logic} 
\subsection{Propositions and Connectives}
Our logic's propositions $\pProp$ are predicates over $\PSpAuth{\monoid}$ that are upwards-closed with respect to the resource order.
The conditioning ($\cond{\dist}{\wsumInx}{\pred(\wsumInx)}$) and update ($\pvs{} P$) modalities were already defined in Section \ref{sec:model}. $\ownM x, \mfble{}, \distAs{X}{\dist}$ and sure propositions ($\sure{\Phi}$) can be defined as follows:
\begin{alignat*}{2}
   \ownM{x'}(x) &\eqdef  x' \resOrder x &\qquad
   (\distAs{X}{\dist})(\fraginstance, \rvRes, \authinstance) &\eqdef X \text{ is distributed as } \dist \text{ in } \fraginstance \\
   (\mfble{\Phi})(\fraginstance, \rvRes, \authinstance) &\eqdef \Phi (\fraginstance, \rvRes, \None) &\qquad
   \sure{\Phi} &\eqdef \distAs{\Phi}{\dirac{\True}}
\end{alignat*}

The remaining primitive connectives $*, \wand, \forall, \exists, \land, \lor, \implies$, pure propositions, and Iris's persistency modality $\always$ for duplicable assertions can all be defined in the standard way. As examples, we show the definitions of $*, \wand$ and $\always$.
\begin{alignat*}{3}
   &(P * Q) (x) &&\eqdef \exists x_1, x_2.~ x_1 \cdot x_2 \mincl x \land P(x_1) \land Q(x_2) \\
   &(P \wand Q) (x) &&\eqdef \forall x'.~\mval(x \cdot x') \land P(x') \implies Q(x \cdot x')  \\
   &(\always P) (x) &&\eqdef P(\mcore{x}) 
\end{alignat*}
Entailment can be defined in the standard way $P \entails Q \eqdef \forall x.~ \mval(x) \implies P(x) \implies Q(x)$.

In the following, we instantiate $\monoid$ with $\Auth(gmap (\Loc \to \Val))$, the resource algebra that is used in standard Iris to model (deterministic) points-to assertions. Our points-to assertion can then be defined as follows.
\[
  L \mapsto V \eqdef \ownM{\munit, (\lambda \outcome.~ \authfrag [L(\outcome) \mapsto V(\outcome)]), \None}
\]

\begin{figure*}[t]
  \begin{proofrules}
  \infer*[lab=distr-subst]{}{
    \distAs{X}{\dist} * \sure{X = Y} \entails
    \distAs{Y}{\dist}
  } \label{rule:distr-subst}

  \infer*[lab=distr-dirac]{}{
    \entails
    \distAs{v}{\dirac{v}}
  } \label{rule:distr-dirac}

  \infer*[lab=own-subst]{}{
    \ownM{\munit,X} * \sure{X = Y}
    \entails
    \ownM{\munit,Y}
  } \label{rule:own-subst}

  \infer*[lab=sure-persistent]{}{
    \sure{X = Y}\entails \always\sure{X = Y}
  } \label{rule:sure-persistent}

    \infer*[lab=c-skolem]{}{
    \cond{\dist}{v}{\exists a.~ \Phi(a, v)}
    \entailEq
    \exists f.~ \cond{\dist}{v}{\Phi(f(v), v)}
    } \label{rule:c-skolem}

  \infer*[lab=c-upd-swap]{}{
    \cond{\dist}{x}{\upd P(x)}
    \entails
    \upd\cond{\dist}{x}{P(x)}
  }

  \infer*[lab=c-bind]{}{
    \cond{\dist}{v}{\distAs{X}{\kappa(v)}}
    \entails
    \distAs{X}{\bind(\dist, \kappa)}
  } \label{rule:c-bind}

  \infer*[lab=c-forall]{}{
    \cond{\dist}{v}{\forall x.~\Phi(x, v)}
    \entails
    \forall x.~\cond{\dist}{v}{\Phi(x, v)}
  }

  \infer*[lab=c-fuse]{}{
    \cond{\dist}{v}{\cond{\kappa(v)}{w}{\Phi(v, w)}}
    \entailEq
    \cond{\dist\fuse \kappa}{(v, w)} {\Phi(v, w)}
  } \label{rule:c-fuse}

  \infer*[lab=c-rv]{}{
    \cond{\dist}{v}{\Phi(v)}
    \entails
    \exists X. \cond{\dist}{v}{\sure{X = v} * \Phi(v)}
  } \label{rule:c-rv}

  \infer*[lab=c-intro]{}{
    \distAs{X}{\dist}
    \entailEq
    \cond{\dist}{v}{\sure{X = v}}
  } \label{rule:c-intro}

    \infer*[lab=c-dirac]{}
  {
    \cond{\dirac{i}}{v}{\Phi(v)}
    \entailEq
    \Phi(i)
  } \label{rule:c-dirac} 
  
  \infer*[lab=c-mono]{
    \forall v \in \supp(\dist).~ \Phi_1(v)
    \entails
    \Phi_2(v)
  } {
    \cond{\dist} {v} {\Phi_1(v)}
    \entails
    \cond{\dist} {v} {\Phi_2(v)}
  } \label{rule:c-mono}

  \infer*[lab=pointsto-str-convex]{}{
    \strconvex{L \mapsto V}
  } \label{rule:pointsto-str-convex}

\infer*[lab=distr-str-convex]{}{
    \strconvex{\distAs{X}{\dist}}
  } \label{rule:distr-str-convex}

  \infer*[lab=distr-frameable]{}{
    \distAs{X}{\dist} \entails \mfble{\distAs{X}{\dist}}
  } \label{rule:distr-frameable}

  \infer*[lab=pointsto-frameable]{}{
    L \mapsto V \entails \mfble{L \mapsto V}
  } \label{rule:pointsto-frameable}

   \infer*[lab=c-frame]{}{
      \mfble P * \cond{\dist}{v}{Q(v)}
      \entails
      \cond{\dist}{v}{(\mfble P * Q(v))}
   } 

  \infer*[lab=sep-frameable]{}{
    \mfble{P} * \mfble{Q} \entailEq \mfble{(P * Q)}
  } \label{rule:sep-frameable}

  \infer*[lab=forall-frameable]{}{
    \mfble{\forall v.~ \Phi(v)} \entailEq \forall v.~ \mfble{\Phi(v)}
  } \label{rule:forall-frameable}

  \infer*[lab=exists-frameable]{}{
    \mfble{\exists v.~ \Phi(v)} \entailEq \exists v.~ \mfble{\Phi(v)}
  } \label{rule:exists-frameable}

\end{proofrules}
$\dist \fuse \kernel \eqdef \lambda(v,w).~\dist(v)\kernel(v,w)$ \qquad\qquad $\strconvex{\Psi} \eqdef \cond{\dist}{v}{\Phi(v) * \Psi}
    \entails
    \Psi * \cond{\dist}{v}{\Phi(v)}$
  \caption{Selected proof rules of \thelogic{}.}
  \label{fig:rules}
\end{figure*}

Selected primitive proof rules are shown in Figure \ref{fig:rules}, a more complete list can be found in \appendixref{sec:all-rules}{A}. \ref{rule:distr-subst} and \ref{rule:own-subst} allow for rewriting with almost-sure equalities. \ref{rule:sure-persistent} states the persistency of almost-sure equalities, allowing to duplicate them anytime. \ref{rule:c-skolem} allows to lift out an existential under conditioning to an existential over a function, \ie Skolemization across the modality. We provide an example of its use in the derivation of \ref{rule:c-exists}. \ref{rule:c-upd-swap} allows replacing one global update with multiple branch-wise ones, as discussed in Section \ref{sec:key:update}. \ref{rule:c-bind} reflects the law of total probability: the distribution of $X$ in all branches can be merged into its global distribution. \ref{rule:c-fuse} allows merging and splitting conditionings: a rule that was shown to be crucial for Bayesian reasoning \cite{bluebell}. We provide an example of its usage in \thelogic~ in Section \ref{sec:random-shuffle} and \appendixref{sec:markov-appendix}{A.1}. \ref{rule:c-rv} makes the random variable that is conditioned on explicit. \ref{rule:c-intro} is the standard rule for introducing conditionings: Ownership of a random variable can be converted into conditioning on that random variable. \ref{rule:c-mono} states the monotonicity of the conditioning modality, allowing one to prove an entailment between two conditioning modalities by case analysis on all branches.

The remaining rules are about getting assertions out of (strong convexity) or into (frameability) the conditioning modality: points-tos (\ref{rule:pointsto-str-convex}) and distributional assertions (\ref{rule:distr-str-convex}) are \textit{strongly convex}, meaning that they can be moved out of a conditioning modality. The converse is frameability, expressed by the frameable modality $\mfble{}$ discussed in Section \ref{sec:key:auth}. All frameable assertions can be moved into conditioning (\ref{rule:c-frame}). Points-tos (\ref{rule:pointsto-frameable}) and distributional assertions (\ref{rule:distr-frameable}) are frameable, and the frameable modality commutes with the separating conjunction (\ref{rule:sep-frameable}) and existential and universal quantification (\ref{rule:exists-frameable}, \ref{rule:forall-frameable}). Note that existentials are frameable but not in general strongly convex since the witness may differ between branches: this is the main difference between the two concepts. 

Handling existentials under conditioning is known to be challenging in general. For the special case of an existential over a random variable, we can \emph{derive} the following general rule within Amaryllis, a rule that several of our examples (Section \ref{sec:eval}) make use of:
\begin{proofrules}
  \infer*[lab=c-exists-rv]{\cond{\dist}{v}{\exists V.~\Phi(v)(V)} \qquad \forall v, V_1, V_2. \sure{V_1 = V_2} * \Phi(v)(V_1) \entails \Phi(v)(V_2)}{\exists V.~\cond{\dist}{v}{\Phi(v)(V)}} \label{rule:c-exists}
\end{proofrules}
\ref{rule:c-exists} allows to pull an existential over a random variable out of a conditioning modality, if the predicate under conditioning $\pred$ is proper with respect to almost-sure equality of the random variable. Note that this side condition is not trivially true since $\pred$ is a meta-level function that takes a random variable $\rvval$ as an argument and may, for example, include a pure assertion $\forall \outcome.~\rvval(\outcome)=0$. However, if $\rvval$ is not used within pure assertions, the side condition is always satisfied.

\ref{rule:c-exists} is derivable from the other rules as follows. First, apply \ref{rule:c-skolem} to the first premise to obtain a  function $f$ satisfying $\cond{\dist}{v}{\Phi(v)(f (v))}$.
Next, use \ref{rule:c-rv} to obtain the random variable $\distAs{X}{\dist}$ that the conditioning is on. 
Now, we can exploit that $\Phi$ does not depend on the probability 0 values of $V$: we choose $V(\outcome)$ by first finding the unique branch $v$ in which $\outcome$ does not have probability 0, and then using the witness $f (v)(\outcome)$ from that branch. 
Formally, we define $V:=\fun \outcome. f (X (\outcome)) \outcome$ and apply \ref{rule:c-mono} to get the goal $\sure{X = v} * \Phi(v)(f (v)) \entails \Phi(v)(\fun \outcome.~f (X (\outcome)) \outcome)$, which follows from the rule's second premise. \qed

\subsection{The Weakest Precondition Modality}

We define the \textit{weakest precondition modality} $\mathsf{wp} : \rv{\Expr} \to (\rv{\Val} \to \pProp)$ using the primitive connectives. This follows the standard structure of an Iris weakest precondition, with the main difference that we are using a big-step instead of a small-step semantics.
	\begin{align*}
		&\wpre \rvexpr{\Ret\rvval. \pred(\rvval)} \eqdef{} \forall \psp, \rvstate. ~
		  \rvexpr, \rvstate~ \text{measurable in} ~ \psp \wand  
			\stateinterp(\psp, \rvstate) \wand   \\
		&\qquad \forall \rvval, \psp', \rvstate'.~\bigstep{\psp}{\fun \outcome.~\rvexpr(\outcome),\rvstate(\outcome)}{\psp'}{\fun \outcome.~(\rvval(\outcome), \rvstate'(\outcome))} \wand
			\pvs \stateinterp(\psp', \rvstate') * \pred(\rvval) 
	\end{align*}
  where the \textit{state interpretation} $\stateinterp$ is defined as 
  $
    \stateinterp(\psp, \rvstate) := \ownM{\mcore{\psp},
      (\lambda \outcome.~ \authfull \rvstate (\outcome)), \psp }
  $
  and the postcondition $\pred$ is required to be proper with respect to almost-sure equality: $\sure{V = V'} * \pred(V) \entails \pred(V')$. 

  In the weakest precondition, we first quantify over all possible current `probabilistic states': A pair $(\psp, \rvstate)$ is a \textit{possible probabilistic state} if (1) both $\rvexpr$ and $\rvstate$ are measurable in $\psp$, and (2) $\psp$ and $\rvstate$ are both \text{consistent} with the `local information', i.e. the assertions in a Hoare triple's precondition. In other words, it is not contradictory to assume that $\psp$ and $\rvstate$ represent the ground truth. Statement (2) is equivalent to assuming the state interpretation.
  
  Next, we quantify over all random value variables and probabilistic states that the current random expression variable might evaluate to. For each option, we require that we can update our local information such that (1) our local information is again consistent with the new probabilistic state $\psp', \rvstate'$, and (2) the postcondition holds on the output random value variable.

  Hoare triples are defined on top of wp:
  $\hoare P \rvexpr Q \eqdef P \entails \wpre \rvexpr Q$.

  \begin{figure*}[t]
\begin{proofrules}
      \infer*[lab=wp-alloc]{}
      {
        \entails \wpre{\Alloc{V}}{\Ret V'. \exists L.~ \sure{V' = L} * L \mapsto V}
      } \label{rule:wp_alloc}

      \infer*[lab=wp-sample]{}{
        0 < p < 1 \entails
        \wpre{\sample(p)}{\Ret V. \distAs{V}{\Ber{p}}}
      } \label{rule:wp-sample}

      \infer*[lab=wp-store]{}{
        L \mapsto V
        \entails
\wpre{\Set{L}{V'}}{L \mapsto V'}
} \label{rule:wp-store}

      \infer*[lab=wp-load]{}{
        L \mapsto V
        \entails
\wpre{\Get{L}}{\Ret V'. L \mapsto V * \sure{V=V'}}
} \label{rule:wp-load}

      \infer*[lab=wp-pure]{
        E \leadsto_{\textit{pure}} E'
      } {
        \wpre{E'}{\Phi} \entails \wpre{E}{\Phi}
      } \label{rule:wp-pure}

      \infer*[lab=wp-as-eq]{}
      {
        \sure{E_1 = E_2} \wand \wpre{E_1}{\Phi} \wand \wpre{E_2}{\Phi}
      } \label{rule:wp-as-eq}

      \infer*[lab=wp-value]{}
      {
        \pvs \Phi(V)
        \entails
        \wpre{V}{\Phi}
      } \label{rule:wp-value}

      \infer*[lab=wp-mono]{}{
        \wpre{E}{\Phi}
        \wand
        (\forall V.~ \Phi (V) \wand \pvs \Psi(V))
        \wand
        \wpre{E}{\Psi}
      } \label{rule:wp-mono}

    \infer*[lab=wp-bind]{}{
      \wpre{E}{\Ret V. \wpre{K[V]}{\Phi}}
      \entails
      \wpre{K[E]}{\Phi}
    } \label{rule:wp-bind}
    \end{proofrules}
    \caption{Selected proof rules about weakest preconditions.}
    \label{fig:wp-rules}
  \end{figure*}

  Proof rules for the weakest precondition can be found in Figure~\ref{fig:wp-rules}. The rules for non-probabilistic constructs are analogous to standard proof rules in Iris, except that they now apply to random variables rather than deterministic expressions. In particular, our points-to assertions behave just like ordinary ones and can also be used in the standard way, abstracting away that the identity of the location may differ under different random outcomes. 
  
  \ref{rule:wp-bind} is the underlying weakest precondition law for \ref{rule:hoare-bind} discussed in Section \ref{sec:key:specs}. Here, $K: \Outcomes \to \Lctx$ is a random evaluation context variable, providing one \lang~ context per outcome. We only allow context-expression decompositions that satisfy $\forall \outcome_1, \outcome_2.~K[\rvexpr](\outcome_1) = K[\rvexpr](\outcome_2) \implies K(\outcome_1) = K(\outcome_2)$ and $\rvexpr(\outcome_1) = \rvexpr(\outcome_2)$, \ie the decomposition must not distinguish between cases that are indistinguishable in the original random expression variable.

  The following adequacy theorem, our main result, states that our WP definition is sound.
\begin{theorem}[Adequacy] 
  If $\expr, \sigma \stdstep \dist$ and
  $\entails \wpre {(\fun\wtv.\expr)} {\Ret V. \distAs{V}{\dist_V}}$, then $\dist_V = \projout{V}{(\dist)}$
\end{theorem}
 \section{Illustrative Examples}
\label{sec:eval}

\subsection{Deriving Uniform Sampling from Bernoulli Sampling}

Our first example illustrates the basic reasoning in \thelogic, especially how our rule \ref{rule:c-wp-swap} enables reasoning about conditional program expressions modularly. The example itself is not novel but meant to serve as a warmup. We verify the following function $\operatorname{u}$, which implements a uniform sampler over the integers $\{0,1,\ldots,n-1\}$ using our Bernoulli sampler, $\sample$.
\begin{equation*}
    \Rec{u}{n} \is
      \If n = 1 then 0 \Else
        \If \sample(\tfrac{1}{n}) then (n - 1) \Else \operatorname{u}(n-1)
  \label{ex:sample-uniform}
\end{equation*}
We show $\hoare{n \in \mathbb{N}_{> 0}}{\operatorname{u}n}{V.~\distAs{V}{\Unif{n}}}$ by induction on $n$. The base case ($n=1$) follows directly from \ref{rule:wp-pure}, \ref{rule:wp-value} and \ref{rule:distr-dirac}.
For the induction step ($n>1$), after beta-reducing $\operatorname{u}n$ with \ref{rule:wp-pure}, we then apply \ref{rule:wp-bind} and \ref{rule:wp-sample} to obtain the proof goal:
\begin{align*}
\distAs{V}{\Ber{1/n}} \entails \wpre{\If V then (n-1) \Else \operatorname{u}(n-1)}{V'.~\distAs{V'}{\Unif n}}
\end{align*}
Next, we would like to condition on $V$ but to do that, we have to transform the left and right sides of the entailment to forms that have the conditioning connective $\cond{\Ber{1/n}}{v}$ at the top-level so that we can apply \ref{rule:c-mono}.
On the left side, we apply \ref{rule:c-intro} to obtain $\cond{\Ber{1/n}}{v}{\sure{V = v}}$.
On the right side, we first apply \ref{rule:wp-mono} and \ref{rule:c-bind} to reduce the postcondition $V'.~\distAs{V'}{\Unif n}$ to $\cond{\Ber{1/n}}{v}{\distAs{V'}{(\text{if } v \text{ then } \dirac{n-1} \text{ else } \Unif{n-1})}}$, obtaining the revised goal:
\begin{align*}
\cond{\Ber{1/n}}{v}{\sure{V = v}} \entails\; &\wpre{\If V then (n-1) \Else \operatorname{u}(n-1)}{\cond{\Ber{1/n}}{v}{\distAs{V'}{(\text{if } v \text{ then } \dirac{n-1} \text{ else } \Unif{n-1})}}}
\end{align*}
This brings us to the key step in the proof: We apply \ref{rule:c-wp-swap} to rewrite the right-hand side, hoisting the $\cond{\Ber{1/n}}{v}$ outside the WP.
\begin{align*}
\cond{\Ber{1/n}}{v}{\sure{V = v}} \entails \cond{\Ber{1/n}}{v}{&\wpre{\If V then (n-1) \Else \operatorname{u}(n-1)}{\distAs{V'}{(\text{if } v \text{ then } \dirac{n-1} \text{ else } \Unif{n-1})}}}
\end{align*}
Next, we apply \ref{rule:c-mono} and use \ref{rule:wp-as-eq} to rewrite with $\sure{V=v}$. A case analysis on $v$ leaves us with two subgoals corresponding to the $\langkw{then}$ and the $\Else$ branches: 
$\wpre{(n-1)}{V'.~ \distAs{V'}{\dirac{n-1}}}$ and $\wpre{\operatorname{u}(n-1)}{V'.~\distAs{V'}{\Unif{n-1}}}$.
These subgoals follow from \ref{rule:distr-dirac} and the induction hypothesis, respectively. \qed

\subsection{Sorted Linked Lists with Random Elements} \label{sec:sll}

Our next example is a standard pointer-based sorted linked list, with three operations: $\operatorname{init}$ (new empty list), $\operatorname{ins}$ (insert a value) and $\operatorname{del}$ (remove a value).
The list is an interesting example for us because it leverages dynamic allocation, which prior DPLs do not support.
Additionally, \emph{sorted} lists interact with randomness in challenging ways. If a sampled value is inserted into a sorted list, the position at which it ends up depends on what samples are drawn for other elements of the list.
Hence, the \emph{structure} of the list is itself randomized.
Despite this entanglement between data and list structure, we are able to prove a strong specification: If each inserted value's distribution is independent of the list's state at the time of insertion, then any value removed from the list is independent of the remaining list's state at the time of removal. 

We start with an unsurprising implementation of the sorted list's three operations in our language.
\begin{align*}
    \operatorname{init} &\is \fun x. \Alloc{\TT} \\
\operatorname{ins} &\is \fun x\, \loc.~
      (\Rec{loop}{\loc'} = \notag
      \Let v = \Get{\loc'} in \notag
       \If v = \TT \lor x \le \Fst{v} then \Alloc{(x, \loc')} \Else \notag \\
      &\quad \Set{\loc'}{(\Fst{v},~ \operatorname{loop}(\Snd{v}))} ;; \loc'
      ) \loc \\
\operatorname{del} &\is \fun x\, \loc.~
      (\Rec{loop}{\loc'} = \notag
      \Let v = \Get{\loc'} in \notag
      \If v = \TT then \loc' \Else \notag\\
      &\quad \If x = \Fst{v} then \Snd{v} \Else \notag
      \Set{\loc'}{(\Fst{v}, \operatorname{loop}(\Snd{v}))} ;; \loc'
      ) \loc
\end{align*}

\newcommand{\ispsll}{\mathsf{psll}}
\newcommand{\issll}{\mathsf{sll}}
\newcommand{\isll}{\mathsf{ll}}
\newcommand{\islls}{\mathsf{lls}}
\newcommand{\sorted}[1]{\mathsf{sorted}(#1)}

To verify our implementation, we will define a representation predicate $\ispsll: \rv{\Loc} \times 2^{\rv{\mathbb{Z}} \times \Dist(\mathbb{Z})} \to \pProp $ for randomized sorted linked lists. Informally, $\ispsll(L, ds)$ holds iff $ds$ is a set of the form $\{(V_1, \mu_1), \ldots, (V_n, \mu_n)\}$, $L$ is a random list containing the random elements $V_1,\ldots,V_n$ in sorted order, and each $V_i$ is distributed as $\mu_i$. (Technically, $L$, $V_i$s are random variables.)
Equipped with this definition, we prove the following specification for the list operations in \thelogic:
\begin{align*}
  \hoare{\EMP}{&\operatorname{init}~ ()}{V.~\exists L.~ \sure{V = L} * \ispsll(L, \emptyset)}  \\
  \{\ispsll(L, ds) * \distAs{X}{\dist}\}~ &
     \operatorname{ins}~X~L~ 
     \{V.~\exists L'.~\sure{V = L'} * \ispsll(L', (X, \dist) \cup ds)\} \\
  \{\ispsll(L, (X, \dist) \cup ds )\} ~ &
    \operatorname{del}~X~L~ \{V.~\exists L'.~ \sure{V = L'} * \ispsll(L', ds) * \distAs{X}{\dist} \}
\end{align*}
The precondition of $\operatorname{ins}$ requires that the inserted element $X$'s distribution be independent of $\ispsll(L, ds)$. Dually, the postcondition of $\operatorname{del}$ states that the removed element's distribution is independent of the rest of the list.

We define $\ispsll$ in three steps. First, we define the standard list representation predicate $\isll: \rv{\Loc} \to \mathit{list}~ \mathbb{Z} \to \pProp$. $\isll(L, zs)$ holds if the \emph{random location $L$} holds the \emph{deterministic elements} of $zs$ in order. 
\begin{align*}
  \isll(L, []) &\eqdef L \mapsto \fun . () \\
  \isll(L, z :: zs) &\eqdef \exists L'.~L \mapsto (\fun \outcome.(z, L'(\outcome))) * \isll(L', zs)
\end{align*}
Next, we define $\issll(L, zs) := \isll(L,zs) * \sorted{zs}$ to add that the lists be sorted. Using $\issll(L, zs)$ we prove standard non-probabilistic specifications for the list operations, e.g., 
\begin{equation*}
  \{\issll(L, zs_1 \append z::zs_2)\} \spac
  \operatorname{del}~z~L \spac
  \{V. \exists L'.~\sure{V = L'} * \issll(L', zs_1 \append zs_2)\}
\end{equation*}

Finally, we use \thelogic's conditioning operator $\condmod$ to lift $\issll$ to the predicate $\ispsll(L, ds)$ that relates a random list $L$ to a set $ds$ of random elements and their (pairwise independent) distributions.
\begin{align*}
  \ispsll(L, ds) \eqdef
    & \cond{\projout{2}{(ds_1)} \otimes \dots \otimes \projout{2}{(ds_n)}}{(v_1, \dots , v_n)}{\exists zs.~(\issll(L, zs) \ast
    \sure{\projout{1}{(ds_1)} = v_1 \land \dots \land \projout{1}{(ds_n)} = v_n} \ast\\
    &\qquad\qquad zs \text{ is a permutation of } v_1, \dots, v_n )}
\end{align*}

The proof of the probabilistic specification of $\operatorname{init}$ is straightforward.
In verifying $\operatorname{del}$, a key step is transforming the postcondition into a conditioning on $\dist \otimes \projout{2}{(ds_1)} \otimes \dots \otimes \projout{2}{(ds_n)}$ using the strong convexity of $\issll$ (proven using \ref{rule:c-exists}).
After that, an application of \ref{rule:c-wp-swap} and \ref{rule:c-mono} leads to a proof goal with deterministic values, which is established using the specification of $\operatorname{del}$.
The proof strategy for $\operatorname{ins}$ is similar. For more details, refer to our Rocq~development~\cite{artifact}. 

To further test modularity of \thelogic, we verified a client for our list. The client iterates from 0 to $i$ and, at each step, randomly either inserts an element sampled from $\Unif{i}$ or does nothing:
\begin{align*}
  &\Rec{rand\_list}{i} = \If i \le 0 then \operatorname{init}\TT \Else \notag\\
      &\quad\If \flip then \operatorname{ins}(\operatorname{u}(i), \operatorname{rand\_list}(i-1)) \Else \operatorname{rand\_list}(i-1)
\end{align*}
We have shown that, conditional on the outcomes of the $\flip$ operations, we obtain a probabilistic sorted linked list with a length corresponding to the number of successful flips:
\begin{align*}
\{\EMP \} & \operatorname{rand\_list}~n \spac \{V.~ \exists L.~ \sure{V=L} \ast \cond{ \dist_{\mathsf{nflips}}}{bs}{(\exists xs.~\ispsll(L,xs) \ast \text{len}(xs) = \text{\#ones}(bs)) }\}
\end{align*}
This postcondition also entails that the elements of the list are conditionally independent. The proof can be found in our Rocq development~\cite{artifact}. \qed

\subsection{Reservoir Sampling} \label{sec:reservoir-sampling}

Reservoir sampling is a family of algorithms for sampling from a long sequence of unknown size while only traversing it once. Concretely, we implement sampling an element from a linked list of unknown length, returning the \emph{location} of the chosen element. 

The following function, $\operatorname{pick}$, traverses the list once, keeping a candidate $\loc_c$ for the sampled location. When reading the $i$-th element from the list, $\loc_c$ is replaced with a pointer to that element with probability $1/i$. After processing the first $i$ nodes, $\loc_c$ is uniformly distributed among them.
\begin{align*}
    \operatorname{pick\_loop} &\is \Rec{loop}{\loc_c~i~\loc} = \notag
      \Let v = \Get{\loc} in
      \If v = \TT then \loc_c \Else \notag \\
      & \qquad \qquad \qquad \qquad \Let \loc_c' = (\If \sample(\tfrac{1}{i}) then \loc \Else \loc_c) in
        \operatorname{loop}~\loc_c'~(i + 1)~(\Snd{v}) \\
    \operatorname{pick} &\is \fun \loc.~ \operatorname{pick\_loop}~\loc~2~(\Snd (\Get{\loc}))
\end{align*}

\noindent For the specification of $\operatorname{pick}$, we need a variant of the $\operatorname{ll}$ predicate that exposes the concrete random location variables:\begin{align*}
  \islls(L, [], []) &\eqdef L \mapsto \fun . () \\
  \islls(L, L' :: Ls, x :: xs) &\eqdef L \mapsto (\fun \outcome.(x, L'(\outcome))) * \islls(L', Ls, xs)
\end{align*}

The specification of $\operatorname{pick}$ should say that the returned location, say $L_p$, is (almost surely) equal to one of the locations in the linked list, and that all locations have equal probability of being returned. Writing this formally is subtle: a triple must hold under every frame and the frame determines the distribution of $L::Ls$, so the returned location's exact distribution varies with the allocation's context. Hence, no triple can specify the output distribution precisely. However, the position of the node of the list that the result points to is frame-independent. Our specification exploits this: It stipulates the existence of a uniform index $V$ that satisfies the almost-sure equality $\sure{L_P = (L :: Ls)[V]}$. 
\begin{align*}
  \{\islls(L, Ls, xs)\} \spac \operatorname{pick}(L) \spac
    \{L_P.~ \islls(L, Ls, xs) \ast \exists V.~ \distAs{V}{\Unif{|xs|}} \ast
      \sure{L_P = (L :: Ls)[V]}\}
\end{align*}

This specification is strong enough to read from, and allocate at, a randomly chosen position as in the following two illustrative clients:
\begin{align*}
  \{\islls(L, Ls, xs)\} \spac & \Fst (\Get(\operatorname{pick}(L))) \spac
    \{V.~ \islls(L, Ls, xs) \ast \distAs{V}{\Unif{xs}}\} \\
  \{\islls(L, Ls, xs)\} \spac & \Let p = \operatorname{pick}(L) in \Set{p}{(v, \Alloc(\Get p))}  \\
    &\qquad \{\cond{\Unif{|xs|}}{k}{\exists L_p.~\islls(L,~ Ls[..k] \append L_p :: Ls[k..],~ xs[..k] \append v::xs[k..])}\}
\end{align*}
Here, the second client inserts a new element $v$ into the list at the location returned by $\operatorname{pick}$.

Existing DPLs cannot handle specifications like the one of $\operatorname{pick}$ since they do not support locations as \emph{first-class values}. In contrast, \thelogic~ treats random location variables just like any other random variable; thus, the logic's full expressive power also applies to them.

For the proof of the example, see our Rocq development \cite{artifact}.

\subsection{Random Shuffle} \label{sec:random-shuffle}

In this section, we verify a program that computes a random shuffle of a linked list by repeatedly deleting a uniformly chosen element from the list and appending it to a new list. This is another example where the shape of memory heavily depends on the random choices made, thus a model that requires heap disjointness across all random outcomes is unable to reason about it.

We implement the shuffle using two standard (non-probabilistic) linked-list operations: $\operatorname{len}~\loc$ returns the length of the list at $\loc$, and $\operatorname{del\_nth}~i~\loc$ removes the node at the (0-based) index $i$, returning a pair of the deleted value and the head location of the resulting list. 
\begin{align*}
    \operatorname{shuffle} &\is \fun \loc.~
      \Let n = \operatorname{len}\loc in
      \Let \mathit{out} = \Alloc\TT in
      \operatorname{shuffle\_loop}~n~\loc~\mathit{out} \\
    \operatorname{shuffle\_loop} &\is \Rec{shuffle\_loop}{n~\mathit{inp}~\mathit{out}} =
      \If n = 0 then \mathit{out} \Else \notag \Let r = \operatorname{del\_nth}~(\operatorname{u}(n))~\mathit{inp} in\\
      &\quad 
      \Let \mathit{out}' = \Alloc(\Fst r, \mathit{out}) in \notag 
      \operatorname{shuffle\_loop}~(n - 1)~(\Snd r)~\mathit{out}'
\end{align*}
Reusing the linked-list predicate $\isll$ from before, we prove that, provided the input list holds pairwise distinct values $xs$, the shuffle produces a list that is uniformly distributed over all permutations:
\begin{align*}
  \{\isll(L, xs)\} \spac \operatorname{shuffle}(L) \spac \{V.~\exists L'.~\sure{V = L'} \ast
    \cond{\Unif{\operatorname{perms}(xs)}}{ws}{\isll(L', ws)}\}
\end{align*}
After the $k$-th iteration of $\operatorname{shuffle\_loop}$, the output list is a sequence of $k$ distinct elements from the input list, and all such sequences are equally probable. Let $\operatorname{k\_perms}~xs~k$ be the set of all sequences of $k$ distinct elements from $xs$. Then,
\begin{align*}
  \forall k \le |xs|.\quad \big\{\,&\cond{\Unif{\operatorname{k\_perms}~xs~k}}{zs}{\isll(L_{\mathit{in}}, xs \setminus zs) \ast \isll(L_{\mathit{out}}, zs)}\,\big\} 
  &\operatorname{shuffle\_loop}~(|xs| - k)~L_{\mathit{in}}~L_{\mathit{out}} \\
  \big\{\,&V.~\exists L'.~\sure{V = L'} \ast \cond{\Unif{\operatorname{perms}(xs)}}{ws}{\isll(L', ws)}\,\big\}
\end{align*}

The proof is by induction on $|xs| - k$. In the induction step, the key steps are the following: The new sample is first transformed into a conditional statement using \ref{rule:c-intro} and moved inside the existing conditioning using \ref{rule:c-frame}. Then, both conditionings are fused into one on the product distribution using \ref{rule:c-fuse}. Finally, the product is transformed into the isomorphic $\Unif{\operatorname{k\_perms} xs~(k+1)}$. The reasoning about the deletion follows the pattern of Example \ref{sec:sll}.

\subsection{Markov Blankets}\label{sec:markov-blankets}
Our last example is a Markov blanket: We sample three values
$x_1, x_2, x_3$ with $x_2, x_3$ being dependent on $x_1, x_2$
respectively, and show that $x_1$ and $x_3$ are independent
\emph{conditional on $x_2$} ($x_2$ blankets or covers $x_1$'s
influence on $x_3$; hence, the term Markov blanket).
The example, shown below, is an adaptation of a key example included in
the \Bluebell\ paper~\cite{bluebell}. The example shows that we inherit Bluebell's ability to do Bayesian reasoning using the conditioning modality, but can obtain a stronger, more modular specification: Our triple does not depend on the concrete implementation of the Markov chains' events, relying only on triples for the functions implementing them.
\begin{equation*}
  \begin{aligned}
    \mathsf{chain}(f,g,h) \eqdef \;
    &\Let x_1 = f\TT in
    \Let x_2 = g(x_1) in 
    \Let x_3 = h(x_2) in
    (x_1, x_2, x_3)
  \end{aligned}
  \label{ex:markov-blanket}
\end{equation*}
We formalize the specification ``$x_1$ and $x_3$ are independent conditional on
$x_2$'' as follows:
\begin{align*}
  &\hoare{\EMP}{\fun . f ()}{V. \distAs{V}{\dist_1}} \implies 
  (\forall x.~\hoare{\EMP}{\fun . g (x)}{V. \distAs{V}{\kernel_2(x)}}) \implies 
  (\forall y.~\hoare{\EMP}{\fun . h (y)}{V. \distAs{V}{\kernel_3(y)}}) \implies  \\
  & \{\EMP\} \spac \mathsf{chain}(f, g, h) \spac \{V.~ \exists V_1, V_2, V_3.~ \sure{V = (V_1, V_2, V_3)} *  
   \exists \dist.~ \cond{\dist}{v}{\left(\sure{V_2 = v} * \exists \dist'.~\distAs{V_1}{\dist'} * \distAs{V_3}{\kernel_3(v)}\right)} \}
\end{align*}

The proof strategy is very similar to Bluebell's and crucially relies on \ref{rule:c-fuse}. The additional modularity is obtained due to the fact that we support \ref{rule:c-wp-swap} without side conditions. More details can be found in \appendixref{sec:markov-appendix}{A.1}. \section{Related and Future Work}
\label{sec:relwork}

We divide probabilistic logics into two main groups:
the \emph{distributional probabilistic logics}~(DPLs) and
the \emph{lifting-based probabilistic logics}~(LPLs).
DPLs provide native support for assertions describing properties of the whole \emph{distribution} over program states
(\eg probabilistic (conditional) independence, which is not directly expressible in existing LPLs).
LPLs instead provide ways to reason about
specific probabilistic properties of interest
(\eg error bounds, expected time complexity, etc.)
within the framework of standard Hoare/separation logics (\eg Iris, WP calculi).

\paragraph{Distributional probabilistic logics}
Among multiple early DPLs, Ellora~\cite{ellora} was the first to introduce the distinction between DPLs and LPLs (which they referred to as ``assertion-based'' and ``expectation-based'' logics, respectively) and to argue for the potential of DPLs as a general-purpose foundation for reasoning about probabilistic programs.
Later, PSL~\cite{psl} pioneered the idea of internalizing independence
via separating conjunction for a discrete probabilistic imperative language
on a fixed set of variables.
The separation of PSL insists on both disjointness of variables (\ie locations)
and probabilistic independence \emph{at the same time}, which (as explained in \Cref{sec:key:dynalloc}) is problematic when considering dynamic memory allocation.
Reasoning under conditioning was then studied in DIBI~\cite{dibi},
which internalized conditioning using a non-commutative connective in addition
to PSL-style separation.

Building on this early work, Lilac~\cite{lilac}
introduced a measure-theoretic model of separation that
internalizes \emph{purely} probabilistic independence of \emph{continuous}
random variables.
Lilac's programs are first-order functional programs with bounded loops,
immutable state, and the ability to sample from continuous distributions.
On top of the new model of separation~(refined in follow-up work~\cite{lilac2}),
Lilac proposed to express reasoning under conditioning
by using a new \emph{conditioning} modality based on disintegrations. Compared to our conditioning modality, there are two main differences: First, Lilac's modality includes the random variable $X$ as an explicit index but leaves its distribution $\dist$ implicit, whereas we include $\dist$ as an explicit index but leave $X$ implicit. Both modality styles are inter-derivable, but some of our rules like \ref{rule:c-bind} benefit from an interface with direct access to $\dist$. Second, Lilac's conditioning modality does not, by default, own $X$---the ownership is lost when introducing conditioning, disabling the reverse direction of \ref{rule:c-intro}. In their examples, they therefore often use the conditioning modality in (non-separating) conjunction with ``own $X$'' to retain ownership.

Later, \Bluebell~\cite{bluebell} proposed a different conditioning modality
(in the context of \emph{discrete} distributions)
which was shown to satisfy very modular entailment rules,
and could encode, explain, and mix both unary and relational concepts
within a unified framework.
The \Bluebell\ logic was the first to show how mutable state could be integrated
with Lilac's notion of separation without conflating disjointness and independence, but its solution is specific to finite variable stores
and does not generalize to dynamic allocation.
Furthermore, \Bluebell\ was the first to propose \ref{rule:c-wp-swap} and recognize its important role in proofs;
their version of the rule, however, requires full ownership of the state to be
applicable, which makes it impossible to apply frame around it, hindering modularity.
Our conditioning modality is modeled after \Bluebell's, and we import \Bluebell's powerful rules for conditioning like \ref{rule:c-skolem}, \ref{rule:c-fuse}, \ref{rule:c-intro}, and our variant of \ref{rule:c-frame}, among others. Our examples in Section \ref{sec:eval} crucially rely on these rules.

\BaSL~\cite{basl} looks at the problem of supporting
the Bayesian updating statement \textbf{\texttt{observe}} in a Lilac-like logic.
Like Lilac, \BaSL\ also lacks a general \ref{rule:c-wp-swap} rule,
and remedies this by providing some specialized rules for conditional sampling. Understanding the conditions under which \ref{rule:c-wp-swap} can be
validated in a model for continuous distributions remains an open problem.

Outcome Separation Logic (OSL)~\cite{osl}
is a probabilistic logic for a language with dynamic allocation,
but where separation is used exclusively for heap disjointness,
not probabilistic independence.

\pcOL~\cite{pcol} extends OSL to integrate (for the first time) invariants and concurrent reasoning into a DPL. \pcOL\ supports almost-surely terminating programs with finite variable stores,
sampling from discrete distributions, and adopts a \Bluebell-style conditioning modality that also enjoys a rather general version of the \ref{rule:c-wp-swap} rule (their \textsc{Split1} and \textsc{Split2} rules).
The proof of soundness, however, relies on the simplifying effect
of adopting a PSL-style separation conflating disjointness and independence,
and thus is fundamentally incompatible with dynamic allocation.
Moreover, our approach based on PFP updates allows us to unify
\textsc{Split1} and \textsc{Split2} and remove some of their side-conditions,
widening their applicability.

Concerning mechanization of DPLs, there is some recent work by Yan et al.~\cite{psl-isabelle} that mechanizes PSL in Isabelle/HOL and applies it to security verification for oblivious algorithms.
However, to our knowledge, \thelogic{} is the first DPL supporting both independence and conditioning that has been mechanized in a proof assistant.

\paragraph{Lifting-based probabilistic logics}
Lifting-based logics for probabilistic programs stick to traditional
assertions over state, but include resources that can
be used to prove certain specific properties of the induced distributions
(\eg expectations, bounds on probability of events).
One subclass of this category are \emph{quantitative logics} where assertions evaluate to a quantity, like probability or expected value, instead of a binary truth value \cite{qsl, wpexp, wpexp2}. This idea enables modular reasoning and leads to automation support, but it is not clear how to apply it to formalize other probabilistic properties like independence.

Most closely related to our work are LPLs based on Iris~\cite{iris},
where assertions are standard separation logic assertions on state,
augmented with ghost state representing some quantity related to the distribution.
Eris~\cite{eris} aims at proving bounds on the probability that the postcondition will be violated.
A distinctive feature of Eris is that the bounds themselves are treated
as ghost resources, called ``error credits''.
Tachis \cite{tachis} uses the same technique to prove bounds on expected runtime, and
Coneris~\cite{coneris} extends it to handle concurrency.

LPLs have also been shown to work well for proving relational properties, \eg equivalence of the distribution induced by two programs.
The pRHL~\cite{prhl} logic pioneered the use of probabilistic couplings
to reduce proofs of probabilistic relational properties to proofs of relational triples with standard assertions on state.
Polaris~\cite{polaris} and Clutch~\cite{clutch} adopt probabilistic couplings to
prove contextual refinements on probabilistic programs in Iris, and
Approxis \cite{approxis} provides a way to prove \emph{approximate} refinements.
\Bluebell~showed how couplings can be derived on top of conditioning
in a relational DPL, which could offer an approach for a future relational extension of \thelogic.

As a result of using standard assertions over state,
the Iris-based LPLs can reuse the Iris infrastructure as is
(including step-indexing and impredicative invariants),
and thus support very rich source languages including higher-order features
and sometimes concurrency.
None of them have internalized probabilistic independence, however;
they have rather been focused on proving specific properties of the program semantics.
As a future direction, it would be interesting to encode these abstractions
(\eg error credits) on top of the \thelogic\ base logic,
for example by deriving variations of our WP that encode
the abstraction-specific invariants.

\paragraph{Nondeterminism and Outcome Logic}
Integrating nondeterminism and random choices in a compositional way
is a longstanding issue, given that conventional solutions have been proven impossible: there is no distributive law combining the probability distribution and powerset monads \cite{no-go}.
A well-known solution, proposed by \textcite{VaraccaW06},
shows that by representing distributions with the more detailed
\emph{indexed valuations},
which record the random choices that lead to an outcome through indexes,
probabilistic choice can be combined with nondeterminism compositionally.
As explained in \cref{sec:key:dynalloc},
the use of an indexed valuations-style model
in \thelogic\ is not motivated by nondeterminism
(of the allocator), but by our need to lift the underlying notion
of separation of state to separation of distributions.
As a byproduct of using indexed valuations,
we can also combine random choices with nondeterminism in the allocator,
and prove triples that are valid
independently of the specific
(non-probabilistic) implementation of allocation one picks.

Demonic Outcome Logic~\cite{demonicol} explored how one can reason about
nondeterminism where the choice can be made according to
unknown probabilistic choices,
by using the convex powerset construction~\cite{VaraccaW06}.
As interesting future work,
one could adapt the operational semantics used by
\thelogic\ so to match the convex sets
of the demonic semantics, and prove triples
which are valid even when the allocator is probabilistic.
This extension might open the opportunity to encode \pcOL-style reasoning
for concurrency in \thelogic.

\paragraph{Advanced language features}
While many DPLs do not support reasoning about almost-surely terminating loops (\ie loops that terminate with probability~1),
standard techniques to prove almost-sure termination
(\eg~\cite{McIverM05,MajumdarS25}) can be integrated in these logics,
as demonstrated by \pcOL~\cite{pcol}.
To be able to do so in \thelogic, we would first need to remove
the finite-support restriction on the probability spaces in the model,
which is the subject of future work.

As we remarked, many Iris-based LPLs can handle advanced features
like impredicative invariants
(and thus can reason about higher-order state and concurrency).
The support for impredicativity and higher-order ghost state~\cite{Jung2016HigherOrder}
hinges on the step-indexing technique~\cite{Appel2001PPC,Ahmed2004Thesis}.
Integrating step-indexing in the model of \thelogic\ is thus an important and challenging
future direction as well.

\newpage
\printbibliography

\ifthenelse{\boolean{appendixincluded}}{
\appendix

\clearpage
\section{Proof Rules} \label{sec:all-rules}
We give a list of proof rules in this section.
The proof rules for conditioning can be found in Figure \ref{fig:app:cond}, for C-frameable in Figure \ref{fig:app:frameable}, for WP in Figure \ref{fig:app:wp} and the inherited rules from Iris in Figure \ref{fig:app:irisrules}

\begin{figure*}[t]
\begin{proofrules}
  \infer*{}{
    \entails
    \distAs{X}{\dist}\wand \sure{X = Y}\wand \distAs{Y}{\dist}
  }

  \infer*{}{
    \entails
    \distAs{(\lambda.~ v)}{\delta(v)}
  }

  \infer*{}{
    \ownM{\munit,X} * \sure{X = Y}
    \entails
    \ownM{\munit,Y}
  }

  \infer*{}{
    \sure{X = Y}\entails \always\sure{X = Y}
  }
\end{proofrules}
\end{figure*}

\begin{figure*}[t]

\begin{proofrules}

  \infer*{}{
    \strconvex{\Psi} \eqdef \cond{\dist}{v}{\Phi(v) * \Psi}
    \entails
    \Psi * \cond{\dist}{v}{\Phi(v)}
  }

  \infer*[lab=c-upd-swap]{}{
    \cond{\dist}{x}{\upd P(x)}
    \entails
    \upd\cond{\dist}{x}{P(x)}
  }

  \infer*[lab=c-mono]{
    \forall v \in \supp(\dist).~ \Phi_1(v)
    \entails
    \Phi_2(v)
  } {
    \cond{\dist} {v} {\Phi_1(v)}
    \entails
    \cond{\dist} {v} {\Phi_2(v)}
  }

  \infer*[lab=c-fuse]{}{
    \cond{\dist}{v}{\cond{\kappa(v)}{w}{\Phi(v, w)}}
    \entailEq
    \cond{\dist\fuse \kappa}{(v, w)} {\Phi(v, w)}
  }

  \infer*[lab=c-forall]{}{
    \cond{\dist}{v}{\forall x.~\Phi(x, v)}
    \entails
    \forall x.~\cond{\dist}{v}{\Phi(x, v)}
  }

  \infer*[lab=c-intro]{}{
    \distAs{X}{\dist}
    \entailEq
    \cond{\dist}{v}{\sure{X = v}}
  }

  \infer*[lab=c-rv]{}{
    \cond{\dist}{v}{\Phi(v)}
    \entails
    \exists X.~\cond{\dist}{v}{\sure{X = v} * \Phi(v)}
  }

  \infer*[lab=c-transf]{
f\text{ injective}\\
    \forall v \in \supp(\dist_1).~ \dist_1(v) = \dist_2(f(v))
  } {
    \cond{\dist_2}{v}{\Phi(v)}
    \entails
    \cond{\dist_1}{v}{\Phi(f(v))}
  } \label{rule:c-transf}

  \infer*[lab=c-skolem]{}{
    \cond{\dist}{v}{\exists a.~ \Phi(a, v)}
    \entailEq
    \exists f.~ \cond{\dist}{v}{\Phi(f(v), v)}
  }

  \infer*[lab=c-bind]{}{
    \cond{\dist}{v}{\distAs{X}{\kappa(v)}}
    \entails
    \distAs{X}{\bind(\dist, \kappa)}
  }

  \infer*[lab=c-false]{
    \Phi(v) \entails \bot\\
    v \in \supp(\dist)
  } {
    \cond{\dist}{v}{\Phi(v)}
    \entails
    \bot
  }

  \infer*[lab=c-dirac]{}
  {
    \cond{\delta(i)}{v}{\Phi(v)}
    \entailEq
    \Phi(i)
  }

  \infer*[lab=c-auth]{
    \forall v\st Q(v) \entails \sure{X=v}
  }{
    \ownM{\authfull g} * \cond{\dist}{v}{Q(v)}
    \entails
    \cond{\dist}{v}{(\ownM{\authfull (g|_{X=v})} * Q(v))}
  }
  
  \infer*[lab=c-pure]{}{
    \supp\dist \subseteq E * \cond{\dist}{v}{\Phi(v)}
    \entailEq
    \cond{\dist}{v}{v \in E * \Phi(v)}
  }

  \infer*[lab=pure-str-convex]{
    \Psi\text{ pure}
  } {
    \strconvex{\Psi}
  }

  \infer*[lab=pointsto-str-convex]{}{
    \strconvex{L \mapsto V}
  }

\infer*[lab=distr-str-convex]{}{
    \strconvex{\distAs{X}{\dist}}
  }

\end{proofrules}
\caption{Proof rules for conditioning}
\label{fig:app:cond}
\end{figure*}

\begin{figure*}[t]
\begin{proofrules}

  \infer*[lab=c-frameable-mono]{
     P \entails Q
  }{
     \mfble P \entails \mfble Q
  }

  \infer*{}{
    \mfble{P} * \cond{\dist}{v}{\Phi(v)} \entails \cond{\dist}{v} {(\mfble{P}) * \Phi(v)}
  }

  \infer*{}{
    \mfble{P} \entails P
  }

  \infer*{}{
    \mfble{\mfble{P}} \entailEq \mfble{P}
  }

  \infer*{}{
    \mfble{\forall v.~ \Phi(v)} \entailEq \forall v.~ \mfble{\Phi(v)}
  }

  \infer*{}{
    \mfble{\exists v.~ \Phi(v)} \entailEq \exists v.~ \mfble{\Phi(v)}
  }

  \infer*{}{
    \plainly P \entails \mfble{P}
  }

  \infer*{}{
    \mfble{P} * \mfble{Q} \entailEq \mfble{(P * Q)}
  }

  \infer*[lab=c-frameable-pers]{}{
    \always P \entails \mfble P
  }

  \infer*{}{
    \distAs X \dist \entails \mfble(\distAs X \dist)
  }

  \infer*{}{
    \ownM{X} \entails \mfble{\ownM{X}}
  }

\end{proofrules}
\caption{Proof rules for C-frameable}
\label{fig:app:frameable}
\end{figure*}

\begin{figure*}[t]
\begin{proofrules}
  \infer*[lab=wp-bind]{}{
      \wpre{E}{\Ret V. \wpre{K[V]}{\Phi}}
      \entails
      \wpre{K[E]}{\Phi}
    }

  \infer*[lab=wp-load]{}{
    L \mapsto V \entails \wpre{\Get{L}}{\Ret X. L \mapsto V * \sure{X = V}}
  }

  \infer*[lab=wp-alloc]{}
  {
    \wpre{\Alloc{V}}{\Ret V'. \exists L.~ \sure{V' = L} * L \mapsto V}
  }

  \infer*[lab=wp-sample]{
    0 < p < 1
  } {
    \wpre{\sample(p)}{\Ret V. \distAs{V}{\Ber{p}}}
  }

  \infer*[lab=wp-store]{}{
    L \mapsto V
    \entails
    \wpre{\Set{L}{V'}}{\Ret V''. \sure{V'' = ()} * L \mapsto V'}
  }

  \infer*[lab=wp-pure]{
    E \leadsto_{\textit{pure}} E'
  } {
    \wpre{E'}{\Phi} \entails \wpre{E}{\Phi}
  }

\infer*[lab=wp-as-eq]{}
  {
    \sure{E_1 = E_2} \wand \wpre{E_1}{\Phi} \wand \wpre{E_2}{\Phi}
  }

  \infer*[lab=wp-value]{}
  {
    \pvs \Phi(V)
    \entails
    \wpre{V}{\Phi}
  }

  \infer*[lab=wp-mono]{}{
    \wpre{E}{\Phi}
    \wand
    (\forall V.~ \Phi (V) \wand \pvs \Psi(V))
    \wand
    \wpre{E}{\Psi}
  }

  \infer*[lab=upd-wp]{}{
    \pvs \wpre{E}{\Phi}
    \entails
    \wpre{E}{\Phi}
  }

  \infer*[lab=wp-upd]{}{
    \wpre{E}{\Ret V. \pvs\Phi(V)}
    \entails
    \wpre{E}{\Phi}
  }

  \infer*[lab=c-wp-swap]{}{
    \cond{\dist}{v}{\Ret V. \wpre{E}{\Phi(v, V)}}
    \entails
    \wpre{E}{\Ret V. \cond{\dist}{v}{\Phi(v, V)}}
  }

\end{proofrules}
\caption{WP rules}
\label{fig:app:wp}
\end{figure*}

\begin{figure}
\begin{proofrules}
\infer*[lab=sep-comm]{}{
    P * Q \entails Q * P
  }

\infer*[lab=sep-assoc]{}{
    (P * Q) * R \entails P * (Q * R)
  }

\infer*[lab=sep-mono]{
    P \entails P' \\ Q \entails Q'
  }{
    P * Q \entails P' * Q'
  }

\infer*[lab=wand-intro]{
    P * Q \entails R
  }{
    P \entails Q \wand R
  }

\infer*[lab=wand-elim]{
    P \entails Q \wand R
  }{
    P * Q \entails R
  }

\infer*[lab=pers-mono]{
   P \entails Q
  }{
     \always P \entails \always Q
  }

\infer*[lab=pers-elim]{}{
  \always P \entails P
  }

\infer*[lab=pers-idemp]{}{
   \always P \entails \always \always P
  }

\infer*[lab=pers-forall]{}{
   \forall a.~\always \Phi(a) \entails \always \forall a.~\Phi(a)
  }

\infer*[lab=pers-exists]{}{
   \exists a.~\always \Phi(a) \entails \always \exists a.~\Phi(a)
  }

\infer*[lab=pers-sep]{}{
  \always P \land Q \entails P * Q
}

\infer*[lab=upd-intro]{}{
    P \entails \pvs P
  }

\infer*[lab=upd-mono]{
    P \entails Q
  }{
    \pvs P \entails \pvs Q
  }

\infer*[lab=upd-trans]{}{
    \pvs \pvs P \entails \pvs P
  }

\infer*[lab=upd-frame]{}{
    P * \pvs Q \entails \pvs P * Q
  }

\end{proofrules}
\caption{Inherited Rules from Iris}
\label{fig:app:irisrules}
\end{figure}

\subsection{Proof of the Markov Blanket Example} \label{sec:markov-appendix}

Here, we sketch the proof for the Markov blanket example in \Cref{sec:markov-blankets}.

The first step of the proof is replacing the postcondition of
$\mathsf{chain}$ with a stronger, easier to use postcondition by
proving the following entailment. The proof of this entailment is the same as in Bluebell's original example.
First, define $\dist_0 \eqdef \dist_1 \fuse \kernel_2$. Following Bayes' theorem, define
$\dist_2 \eqdef \lambda v_2.~\sum_{v_1} \dist_0(v_1, v_2)$ and
$\kernel_1 v_2 \eqdef \lambda v_1.~\frac{\dist_0(v_1, v_2)}{\dist_2(v_2)}$ such that we can invert the conditional probability: $\dist_1 \fuse \kernel_2(v_1, v_2) = \dist_2 \fuse \kernel_1 (v_2, v_1)$.
We use this equality together with \ref{rule:c-transf} and derive the following:
\begin{align*}
  &\cond{\dist_1}{v_1}{\sure{V_1 = v_1} * \cond{\kernel_2(v_1)}{v_2}{\sure{V_2 = v_2} * \distAs{V_3}{\kernel_3(v_2)}}}\\
\entails~&\cond{\dist_1}{v_1}{\cond{\kernel_2(v_1)}{v_2}{\sure{V_1 = v_1} * \sure{V_2 = v_2} * \distAs{V_3}{\kernel_3(v_2)}}} && \ref{rule:c-frame}\\
\entails~&\cond{\dist_1\fuse \kernel_2}{(v_1, v_2)}{{\sure{V_1 = v_1} * \sure{V_2 = v_2} * \distAs{V_3}{\kernel_3(v_2)}}} && \ref{rule:c-fuse}\\
\entails~&\cond{\dist\fuse \kernel}{(v_2, v_1)}{{\sure{V_1 = v_1} * \sure{V_2 = v_2} * \distAs{V_3}{\kernel_3(v_2)}}} && \ref{rule:c-transf}\\
\entails~&\cond{\dist}{v_2}{\cond{\kernel(v_2)}{v_1}{\sure{V_1 = v_1} * \sure{V_2 = v_2} * \distAs{V_3}{\kernel_3(v_2)}}} && \ref{rule:c-fuse}\\
\entails~&\cond{\dist}{v_2}{\sure{V_2 = v_2} * \left(\cond{\kernel(v_2)}{v_1}{\sure{V_1 = v_1}}\right) * \distAs{V_3}{\kernel_3(v_2)}} && \ref{rule:distr-str-convex}\\
\entails~&\cond{\dist}{v_2}{\sure{V_2 = v_2} * \distAs{V_1}{\kernel(v_2)} * \distAs{V_3}{\kernel_3(v_2)}} && \ref{rule:c-intro}\\
\entails~&\exists \dist\,\kernel.~ \cond{\dist}{v}{\sure{V_2 = v} * \distAs{V_1}{\kernel(v)} * \distAs{V_3}{\kernel_3(v)}}\\
\entails~&\exists \dist.~ \cond{\dist}{v}{\sure{V_2 = v} * \exists \dist'\!.~\distAs{V_1}{\dist'} * \distAs{V_3}{\kernel_3(v)}} && \ref{rule:c-mono}
\end{align*}

Using \ref{rule:wp-mono}, our proof goal becomes 
\begin{align*}
  \{\EMP\} \spac & \mathsf{chain}(f, g, h) \spac \{V.~ \exists V_1, V_2, V_3, \sure{V = (V_1, V_2, V_3)} * & \\
  & \cond{\dist_1}{v_1}{\left(\sure{V_1 = v_1} * \cond{\kernel_2(v_1)}{v_2}{\left(\sure{V_2 = v_2} * \distAs{V_3}{\kernel_3(v_2)}\right)}\right)} \}
\end{align*}
To prove this, we first apply $f$'s Hoare triple using \ref{rule:wp-bind} to obtain a random variable $\distAs{V_1}{\dist_1}$ that is stored in $x_1$.
Next, to use $g$'s Hoare triple, we must condition on $V_1$ after applying \ref{rule:c-wp-swap} to the goal.
However, we cannot immediately apply \ref{rule:c-wp-swap} since the conditioning $\cond{\dist_1}{v_1}{(\ldots)}$ in the goal is under an existential quantifier.
We commute the conditioning and existential quantification using \ref{rule:c-exists}. Afterwards, we can apply \ref{rule:c-wp-swap} to condition on $V_1$, and \ref{rule:wp-bind} to use the Hoare triple for $g$. To complete the proof, we repeat this pattern once more for~$h$. \qed

This example cannot be verified in existing DPLs for different reasons.
\Bluebell, due to its \textsc{c-wp-swap} rule which must be used to
verify the three function calls, requires ownership of every variable in
the precondition of the triple,
thus prohibiting framing of resources around
the final $\mathsf{chain}$ triple, hindering its reuse.
All other DPLs conflate probabilistic independence and disjointness and,
therefore, they cannot express assertions like $\mathsf{chain}$'s
postcondition, which mentions all three variables $V_1,V_2,V_3$
on both sides of a separating conjunction.

\newcommand*\vals{\mathsf{val}}
\newcommand*\sigmaintersec{\sigma\_\mathsf{intersec}}
\newcommand*\nullset{\mathcal N}
\newcommand*\isundet{\mathsf{is\_undeterminate}}
\newcommand*\PSpU{\mathsf{PSpU}}
\newcommand*\namespace{N}

\section{Dynamic Domain Extensions with $\PSpD$} \label{sec:full-model}

In this section, we present our finite representation of probability spaces over $\Rid \to \Bool$. We use Bluebell's $\PSpRA_\Omega$ but do not choose a fixed $\Omega$, but instead dynamically extend $\Omega$ on the fly as we encounter new randomness IDs. All randomness IDs that are not mentioned in the current probability space are considered to have an unknown distribution. That way, elements of the resource algebra can still represent the infinite probability space that we are interested in, while actually being finite.

The resource algebra $\PSpD$ explicitly tracks which RIDs are mentioned in the probability space.
Its type is therefore $\Rid \times \PSpRA$.
The \textit{domain extension operation} $\extPfn$ extends an element $(\rids, \salg, \dist) \in \PSpD$ with a set of new RIDs $\rids'$, without adding any information about the distribution of those RIDs.
This operation can be seen as going from a resource algebra element where the RIDs in $\rids'$ are implicitly unknown to an equivalent resource algebra element where they are explicitly mentioned but still unknown.
\begin{align*} 
  &\extPfn((\rids, \salg, \dist), \rids') :=
  \complete (\{ \{m_1 \cup m_2 \mid m_1 \in E \land m_2 : \rids' \rightarrow \mathbb{B} \} \mid E \in \salg \} , \dist_{ext}) \\
  &\dist_{ext} (E) := \dist (\{m : \rids \rightarrow \mathbb{B} |~ \exists m' \in E, m \mincl m'\})
\end{align*}

The function $\complete$ calculates the completion of the probability space.
We can use the domain extension operation to define an equivalence relation on this resource algebra:
\begin{align*}
&(\rids_1, \salg_1, \dist_1) \equiv (\rids_2, \salg_2, \dist_2) :=
\extPfn((\rids_1, \salg_1, \dist_1), \rids_2 \setminus \rids_1) = 
  \extPfn((\rids_2, \salg_2, \dist_2), \rids_1 \setminus \rids_2)
\end{align*}

The resource composition is defined as follows:
\begin{align*} 
  &(\rids_1, \salg_1, \dist_1) \cdot (\rids_2, \salg_2, \dist_2) := 
  (\rids_1 \cup \rids_2,~ 
  \extPfn((\rids_1, \salg_1, \dist_1), \rids_2 \setminus \rids_1)~ \cdot_{PSp}~ 
  \extPfn((\rids_2, \salg_2, \dist_2), \rids_1 \setminus \rids_2))
\end{align*}

The persistent core of a resource algebra element is the \textit{sure} part of the probability space: it consists of all probability 0 and 1 events.
  $$ | (\rids, \salg, \dist) | = (\rids, \{E \in \salg \mid \dist(E) = 0 \lor \dist(E) = 1\}, \dist )$$

\paragraph{Weighted Sum} Next, we'll define the weighted sum for this resource algebra. We would like the weighted sum to represent \textit{conditioning}, thus we require that the decomposition given by the branches of the weighted sum is induced by a random variable (the random variable we are conditioning on). This is equivalent to requiring the support of the branches to be pairwise disjoint.
\begin{align*}
&\hasrv(\dist, \kernel) :=
\forall \rho.~ \forall i~ j \in supp~ \dist.~ \isposs(\rho, \kernel(i)) \rightarrow \isposs(\rho, \kernel(j)) \rightarrow i = j
\end{align*}
where 
$\isposs(\rho, (\salg, \dist)) := \exists m \in supp (\salg, \dist).~\forall \rid \in \dom~(m).~m[\rid] = \rho(\rid) $. \\

In the weighted sum for this resource algebra, an event is measurable if it is measurable in all branches. Its probability is the sum of the probabilities in the branches.
\begin{align*} 
  \sigmaintersec(\dist, \kernel) &:= \bigcap \pi_\sigma(\extPfn(\kernel(i), \bigcup_{i \in \supp(\dist)} \rids (\kernel (i))))
\end{align*}
\begin{align*}
  wsum(\dist,\kernel) &:= bind(\dist, \lambda i.~ \extPfn (\kernel(i),~ \bigcup_{i \in \supp(\dist)} \rids (\kernel(i)) )_{|\sigmaintersec(\dist, \kernel)} ) \\
  & \qquad \qquad \text{if } \hasrv(\dist, \kernel) \land \forall i \in \supp(\dist).~ \mval (\kappa(i))
\end{align*}

$$ \pi_\dist(bind(\dist, \kernel)) (X) = \sum_{v \in supp(\dist)} \dist (v) \cdot \pi_\dist(\kernel(v)) (X) $$

 }{}

\end{document}